\documentclass[eqsecnum,amsmath,onecolumn,notitlepage,nofootinbib,superscriptaddress]{article}
\usepackage{custom}
\usepackage{xcolor}

\pdfoutput=1

\begin{document}

\title{Uniformizing Lee-Yang Singularities }
\author[1]{G\"{o}k\c{c}e Ba\c{s}ar\thanks{gbasar@unc.edu}}
\author[2]{Gerald V. Dunne\thanks{gerald.dunne@uconn.edu}}
\author[1]{Zelong Yin\thanks{zelong@live.unc.edu}}

\affil[1]{Department of Physics and Astronomy, University of North Carolina, Chapel Hill, NC 27599}
\affil[2]{Physics Department, University of Connecticut, Storrs, CT 06269-3046}

\date{}
\maketitle

\begin{abstract}
Motivated by the search for the QCD critical point, we discuss how to obtain the singular behavior of a thermodynamic system near a critical point, namely the Lee-Yang singularities, from a limited amount of local data generated in a different region of the phase diagram. We show that by using a limited number of Taylor series coefficients, it is possible to reconstruct the equation of state past the radius of convergence, in particular in the critical region. Furthermore we also show that it is possible to extend this reconstruction to go from a crossover region to the first-order transition region in the phase diagram, using a uniformizing map to pass between Riemann sheets.
 We illustrate these ideas via the Chiral Random Matrix Model and the Ising Model.

\end{abstract}

\section{Introduction}

Mapping the phase structure of QCD plays an important role in understanding the structure of matter in extreme environments, both theoretically and experimentally. A particularly important aspect of this endeavor is the ongoing search for the QCD critical point, a singular point in the phase diagram where a smooth crossover between the hadronic phase and the quark-gluon plasma phase turns into a first-order transition. It is one of the major motivations of the Beam Energy Scan program at the Relativistic Heavy-Ion Collider as well as future heavy-ion facilities \cite{Bzdak:2019pkr}. Quantitative theoretical knowledge of the QCD critical point and the equation of state in its vicinity is crucial for the experiments to identify critical point signatures \cite{best}. 

The fermion sign problem complicates the use of lattice QCD to explore the phase diagram at non-zero baryon chemical potential, $\mu_B$, where the critical point is conjectured to exist. Without direct access to the phase diagram, typical methods to extract information on the phase diagram at finite density include Taylor expanding around $\mu_B=0$ \cite{Bazavov:2017dus}, simulating QCD with an imaginary chemical potential where there is no sign problem and analytically continuing to real values \cite{Ratti:2018ksb}, and resummation methods some of which incorporate both \cite{Borsanyi:2021sxv, mondal2021lattice, Mukherjee:2021tyg,Dimopoulos:2021vrk,Singh:2021pog}. 

At the same time, even without a direct calculation of the critical point it is still possible to predict some of its properties, if it exists. This is due to the fact that based on general symmetry arguments, the QCD critical point is in the same static universality class as the three dimensional Ising model  \cite{Stephanov:2004wx}. Universality essentially relates the singular contribution to the QCD equation of state to the equation of state of the Ising model in the vicinity of the critical point. This fixes the critical exponents which determine how certain thermodynamic functions diverge. However, the precise form of this relation is not determined by universality. The relationship between the Ising parameters (namely the reduced temperature $r$ and the magnetic field $h$) and those of QCD (the temperature $T$ and chemical potential $\mu$) is not determined by universality and has to be extracted directly from QCD \cite{best}. Likewise, the location of the critical point is also a non-universal quantity. 

In this paper we tackle these issues from the perspective of series expansions: given a finite-order series expansion around $\mu=0$, we describe improved methods for extracting physical information regarding the singularities of the equation of state in the vicinity of the critical point. In particular we show that with a suitably chosen resummation scheme it is possible to: i) extract the location of the nearest complex singularities, the Lee-Yang edge singularities, which can be used to determine the location of the critical point and constrain the singular contribution to the equation of state; and ii) analytically continue the equation of state across different Riemann sheets in a way that relates the high temperature crossover region to the low temperature first-order transition region. These ideas follow closely Ref. \cite{PhysRevLett.127.171603}, whose mathematical foundations can be found in Refs. \cite{Costin:2020pcj,Costin:2021bay}. 

The paper is organized as follows. In Section \ref{sec:rmm} we briefly summarize the relevant properties of the Chiral Random Matrix Model model that we use to motivate and illustrate our framework. In Section \ref{sec:conf_pade} we explain how to construct the Lee-Yang edge singularities from a series expansion by using the conformal-Pad\'e method, and further to extract the critical point and constrain the equation of state. In Section \ref{sec:uniform} we focus on a different problem and explain how the equation of state can be analytically continued across different Riemann sheets using a different resummation scheme which we call ``uniformized-Pad\'e". In the conclusions we briefly discuss the outlook for future extensions. 

\section{The chiral random matrix model}
\label{sec:rmm}
We introduce the ideas that we develop in this paper via the Chiral Random Matrix Model \cite{Halasz_1998}, which shares some of the key properties with the conjectured QCD phase diagram, such as chiral symmetry breaking and chiral restoration at large chemical potential, as well as the existence of a critical point along this transition curve. In this section we review some of its known properties that will be relevant in our analysis. Readers who are familiar with the model can skip this section. 

\begin{figure}[h]
\center
\includegraphics[scale=0.6]{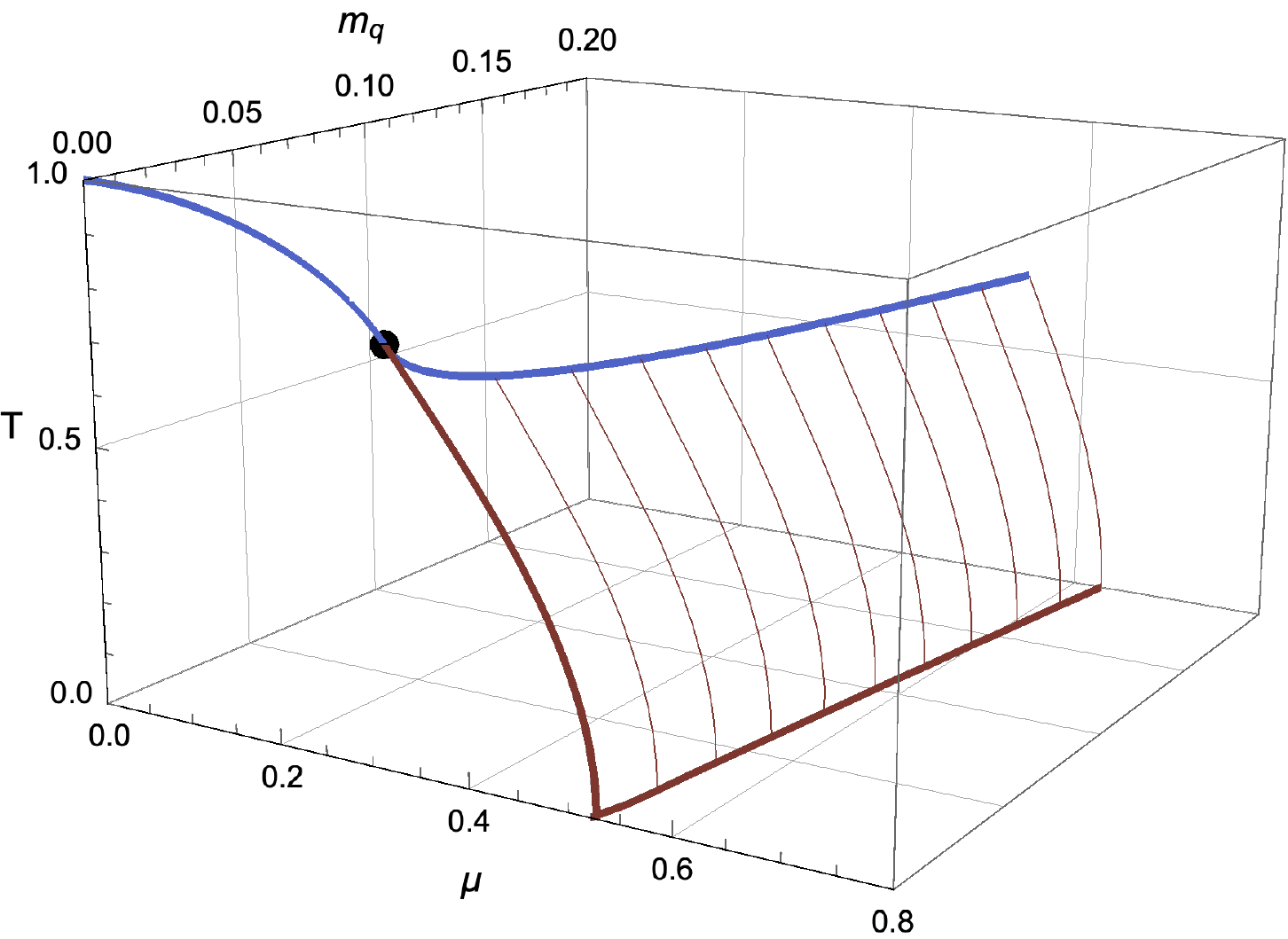}
\caption{The phase diagram of the Chiral Random Matrix Model. Red and blue curves represent the first and second-order transitions and the black dot represents the tri-critical point where the first and second order transitions meet at $m_q$=0. For nonzero values of $m_q$ the tri-critical point turns into a second order critical point with mean field exponents parameterized by $m_q$, $(T_c(m_q),\mu_c(m_q))$. }
\label{fig:rmm-pd}
\end{figure}

The Chiral Random Matrix Model is a toy model for QCD where the matrix elements of the Dirac matrix are replaced by Gaussian random variables.
Its partition function for $N_f$ number of fermions is
\bea
Z(T,\mu)&=&\int {\cal D}\Phi e^{-N {\rm Tr} (\Phi\Phi^\dagger)}
\prod_{f=1}^{N_f}{\rm det}^{N/2}\left(\begin{matrix}\Phi+m_f & \mu+ i T \\ \mu+ i T &\Phi^\dagger+m_f \end{matrix}\right)
{\rm det}^{N/2}\left(\begin{matrix}\Phi+m_f & \mu- i T \\ \mu- i T &\Phi^\dagger+m_f \end{matrix}\right)
\label{eq:ZfiniteN}
\ea
where the integration is over all $N\times N$ complex matrices $\Phi$. We work with $N_f=1$ for the rest of the paper and denote the quark mass as $m_q$. In the $N\rightarrow\infty$ limit the path integral is saturated by the saddle point where the matrix $\Phi$ is proportional to the unit matrix, $\Phi=\phi\times{\mathbb 1}_{N\times N}$ with $\phi\in{\mathbb R}$:
\bea
\lim_{N\rightarrow \infty}\frac{1}{N}\log Z(T,\mu)&=&-\min_{\phi} \Omega (T,\mu,\phi)
\\
\Omega(T,\mu,\phi)&=&\phi^2-\frac12\log\left[ \left( (\phi+m_q)^2-(\mu+i T)^2\right) \left( (\phi+m_q)^2-(\mu-i T)^2\right) \right] \,.
\ea
As usual the minimization of the free energy determines the equation of state where the pressure is identified as $p(T,\mu)=-\min_{\phi} \Omega (T,\mu,\phi)$. The phase diagram of this Chiral Random Matrix Model is shown in Fig. \ref{fig:rmm-pd}. For $m_q=0$ and small values of $T$ and $\mu$, the ground state is given by $\phi\neq0$, which breaks chiral symmetry. At larger values of $T$ and/or $\mu$, chiral symmetry is restored via a second or first order transition, as shown in the figure as blue and red lines respectively. For a fixed $m_q>0$, the chiral symmetry is explicitly broken but the remnant of the second order transition still persists as a rapid crossover. For lower temperatures this crossover turns into a first order transition (red lines in the figure). The point where the crossover turns into a first order transition is a second order critical point. As a function of $m_q$ the critical point follows a trajectory in the three dimensional $T,\mu,m_q$ space, as shown in Figure \ref{fig:rmm-pd}.

We are specifically interested in the physics near the critical point where
\bea
\frac{\partial }{\partial
\phi}\Omega(T,\mu,\phi)=\frac{\partial^2 }{\partial
\phi^2}\Omega(T,\mu,\phi)=0\,.
\label{eq:critical}
\ea
Let us denote the point where a real solution exists as $T=T_c, \mu=\mu_c$, and $\phi=\phi_c$.\footnote{As we will explain further, as a particular property of the mean field limit, $T_c$ corresponds to the largest temperature such a solution exists.} The susceptibility and the heat capacity diverge as power laws at the critical point. Furthermore, the singular part of the equation of state in the vicinity of the critical point is essentially the same as that of the Ising model in the mean field limit, since the Chiral Random Matrix Model and the Ising model are in the same static universality class. The same holds for QCD as well, but in this case it is the three dimensional Ising model as opposed  to the Ising model in the mean field limit. 

In the continuum limit and $d$-dimensions, the Ising model is described by a scalar field with the action
\bea
S=\int d^dx \left[\frac12 (\partial_\mu \phi)^2+\frac12 r_0 \phi^2+\frac14 u_0\phi^4 \right]
\label{eq:action}
\ea
In $d\leq4$ dimensions $r_0$ and $u_0$ flow to the Gaussian fixed point where $u_0$ becomes arbitrarily small and the theory can be described by mean field theory. In mean field theory, the equation of state follows from minimizing the effective potential 
\bea
\Omega_{I}(M)= -h M + \frac12 r M^2+\frac14 M^4 
\ea
where $M$ is the average magnetization $M=\langle{\phi\rangle}$. For convenience we have re-scaled $\phi$ to set the coefficient of the quartic term to $1/4$. The equation of state is obtained by minimizing the effective potential:
\bea
\frac{\partial \Omega_{I}(M)}{\partial M}=-h+ rM+M^3=0\,.
\label{eq:eos}
\ea
The critical point is located at $\partial \Omega_{I}(M)/\partial M=\partial^2\Omega_{I}(M)/\partial_M^2=0$, which corresponds to $r=h=0$. For fixed $r>0$, the magnetization smoothly crosses over from $M<0$ to $M>0$ as the magnetic field varies from $h<0$ to $h>0$. For $r<0$ this transition is first order. As is typical in first order transitions for $r<0$ and $h<0$, in addition to the global minimum state with $M<0$ there is also a meta-stable state with $M>0$. The point in the phase diagram (for fixed $r$) where the meta-stable state ceases to exist is the spinodal point.  
The equation of state \eqref{eq:eos} is invariant under scaling $r\rightarrow \lambda r$, $h\rightarrow \lambda^{3/2}h$, $M\rightarrow \lambda^{1/2}M$. Therefore it is convenient to express the minimization in terms of the scaling variables, $w:=h r^{-3/2}$ and $z:=M r^{-1/2}$, as:
\bea
w=z+z^3\,.
\label{eq:eos_zw}
\ea

\begin{figure}[h]
\center
\includegraphics[scale=0.6]{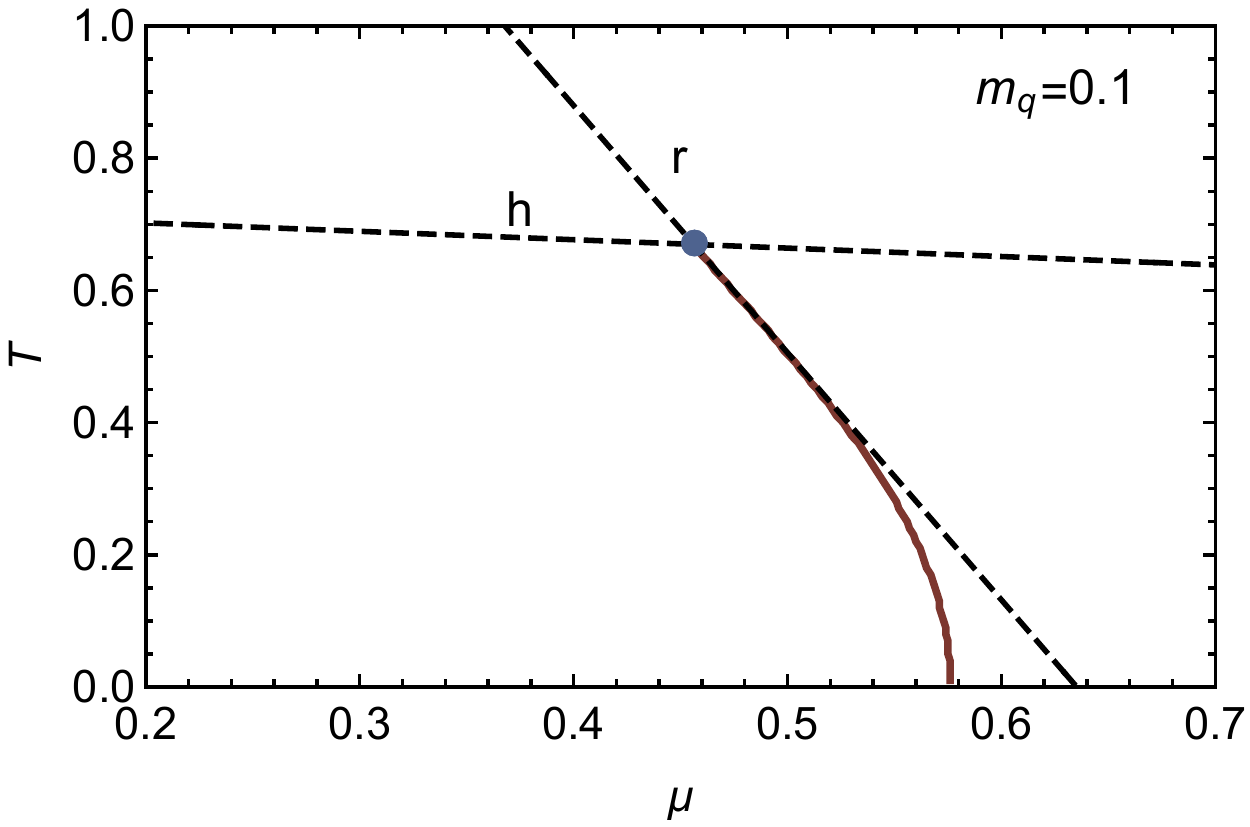}
\caption{The Ising model parameters mapped to the $(T,\mu)$ plane of the chiral random matrix model. The slopes of the $r$ and $h$ axes are $\tan \alpha_1\equiv\hmu/\hT$ and $\tan \alpha_2\equiv \rmu/\rT$, respectively. For $m_q=0.1$ the critical point is at $T_c\approx0.67$ and $\mu_c\approx 0.46$. }
\label{fig:pd}
\end{figure}

In the vicinity of the critical point, the equation of state of the Chiral Random Matrix model can be mapped to that of the Ising model. Let us first expand the temperature and the chemical potential around the critical point as $\Delta T=T-T_c$, $\Delta \mu=\mu-\mu_c$. They can be mapped into the the Ising variables $r$ and $h$ linearly as 
\bea
\label{eq:params}
h=\hT \Delta T+\hmu \Delta \mu,\quad r=\rT \Delta T+\rmu \Delta \mu
\ea
To see how this mapping works let us follow the steps in \cite{Pradeep:2019ccv} and expand $\Omega(\phi)$ around the critical point 
\bea
\label{eq:Omega-exp}
\Omega(\phi)=\Omega(\phi_c)+f_1(T,\mu)\dphi+f_2(T,\mu) \dphi^2+\dots
\ea
where $\dphi=\phi-\phi_c$. Further expanding $f_i$ around $T_c,\mu_c$ and eliminating the $\dphi^3$ term via shifting $\dphi$ by a constant we have 
\bea
\Omega(\phi)=\Omega_0-\hb \dphi+\frac\rb2 \dphi^2+\frac{u}{4}\dphi^4+vu\dphi^5+\ord(\dphi^6)
\ea
where the $\dphi^5$ term is necessary to take into account the correction to scaling that is present in Eq. \eqref{eq:params} since $h$ and $r$ have scale differently ($h\sim r^{3/2}$). This quintic term can be eliminated via $\dphi\rightarrow \dphi+v(\rb/u-\dphi^2)$ to leading order in $r$. Finally rescaling the field so that the coefficient of the quadratic term is $1/4$ leads to the Ising form 
\bea
\Omega(\phi)=\Omega_0-h \dphi+ \frac{r}{2}\dphi^2+\frac14\dphi^4+\dots
\ea
Here $h=u^{-1/4}\hb$, and $r=u^{-1/2}(\rb+2 v \hb)$ in leading order in $\rb$. Since $\rb$ and $\hb$ are obtained from expanding $f_1$ and $f_2$ to linear order in $\Delta T$ and $\Delta \mu$, from these expressions we can read off the mapping parameters $(\hT,\hmu,\rT,\rmu)$. For example, for $m_q=0.1$ we obtain $(\hmu,\hT,\rmu,\rT)\approx(-1.692, -0.452, 0.315, 2.481)$ with $\tan \alpha_1\equiv \hmu/\hT\approx3.739$ and $\tan \alpha_2\equiv\rmu/\rT\approx0.127$ (see Fig. \ref{fig:pd}).

\subsection{Lee-Yang edge singularities}

For $m_q>0$, the critical point \eqref{eq:critical} is a singular point in the phase diagram. However, in general, for $T\neq T_c$ this condition is satisfied for a pair of complex conjugate values $\mu$. These points correspond to the celebrated Lee-Yang (LY) edge singularities \cite{Lee:1952ig} which will denote as $\mu_{LY}(T)$. As $T$ approaches $T_c$ from above they pinch the real axis such that $\mu_{LY}(T_c)=\mu_c$. 

More generally, the finite $N$ partition function, \eqref{eq:ZfiniteN} is a polynomial in $\mu$ of order $2N\times N_f$, which therefore has $2N\times N_f$ zeros. As $N\rightarrow \infty$ they form branch cuts and $\mu_{LY}(T)$ are the branch points associated with these cuts \cite{Halasz_1997,Stephanov:2006dn}. 

In the context of the critical point, we would like to analyze $\mu_{LY}(T)$ in the vicinity of the critical point. Using universality, we can turn to the Ising model where the Lee-Yang singularity is simply $dw/dz=0$, which from Eq. \eqref{eq:eos_zw} leads to $w_{LY}=\pm i 2/(3\sqrt{3})$.\footnote{Beyond the mean field limit this value has been computed in Ref. \cite{Mukherjee:2019eou} for the three dimensional $O(N)$ symmetric model.} The LY singularity can be viewed as a critical point and in its vicinity the equation of state behaves as
\bea
z-z_{LY}\propto (w-w_{LY})^{\sigma_{LY}}
\label{eq:w_LY_eos_}
\ea
where $\sigma_{LY}=1/2$ is the associated critical exponent. Beyond the mean field limit, the critical equation of state in the vicinity of the LY singularity is described by the $\phi^3$ theory with pure imaginary coupling \cite{Fisher:1978pf} with $\sigma\approx0.074-0.085$. 
Using the mapping back to the random matrix model, Eq. \eqref{eq:params}, we obtain
\bea
\mu_{LY}(T)\approx \mu_c+K_1(T-T_c)\pm i K_2 (T-T_c)^{3/2}
\quad\text{where}\quad
K_1 = -\frac\hT\hmu,\quad K_2= \frac{2}{3\sqrt{3}} \frac{\rmu^{3/2}}{\hmu}\left(\frac\rT\rmu-\frac\hT\hmu \right)^{3/2}\,.
\label{eq:lytraj}
\ea
It is worth noting that $\mu_{LY}(T)$ is real for $T<T_c$. In this regime they correspond to the location of the spinodal point. The fact that they lie on the real axis for $T<T_c$ is an artifact of the mean field limit. In general, the sub-leading term in Eq. \eqref{eq:lytraj} is proportional to $(T-T_c)^{\bd}$ where $\beta$ and $\delta$ are the usual critical exponents. In mean field theory $\beta\delta=3/2$ and the sub-leading term becomes real for $T<T_c$. Beyond mean field this is not the case, which has interesting consequences for the spinodal singularities \cite{An:2016lni,An:2017brc,An:2017rfa}. 

The LY trajectory given in Eq. \eqref{eq:lytraj} is the main starting point of our analysis. Our strategy in the next section will be to reconstruct this expansion near $T_c$ from a truncated series expansion of the equation of state. Then one can obtain the location of $T_c$, $\mu_c$, as well as $K_1$ and $K_2$ which contain the mapping parameters to the Ising model. More generally, the LY singularities in the context of QCD critical point have been discussed in, for example, \cite{Damgaard:1993df,Ejiri:2005ts,Stephanov:2006dn,WAKAYAMA2019227,Mukherjee:2019eou,Connelly:2020pno,Schmidt:2021pey}.

\section{The conformal map and the Lee-Yang trajectory}
\label{sec:conf_pade}

In this section we explain how to determine the location of the LY singularities with high precision
in the practical situation where we only have access to approximate information about the equation of state. Furthermore, this approximate information is typically 
computed in a region away from the critical region, and yet we are most interested in probing the vicinity of the critical point (see for example \cite{Fisher:1974series}). Due to the fermion sign problem, most commonly the region we have access to is around $\mu=0$ (see \cite{Ratti:2018ksb} for a recent review), and the information we have is typically a local Taylor expansion around this point. For concreteness let us focus on the susceptibility,
\bea
\chi(T,\mu)=\frac{\partial^2 p(T,\mu)}{\partial\mu^2}\approx\sum_{n=0}^{N} c_{n}(T)\mu^{2n}\,.
\label{eq:eos_Taylor}
\ea
The natural expansion parameter is $\mu^2$, similar to QCD in which case is due to the charge conjugation symmetry. Even though this is a \textit{local} expansion around $\mu=0$, it contains \textit{global} information, including especially the singular behavior around $\mu=\mu_{LY}$. This information is encoded in the coefficients $c_n(T)$, and our task is to decode it as efficiently and precisely as possible. Optimizing this decoding procedure is important as in many cases we only have access to the first few terms in the local expansion. In this section we introduce an efficient framework that not improves the approximation to the LY singularity compared to other methods, but also provides an accurate approximation to the equation of state in the critical region. 
The ideas we pursue here are built upon techniques developed in \cite{Costin:2020pcj,Costin:2021bay}, and which have recently been applied to the Gross-Neveu model in Ref. \cite{PhysRevLett.127.171603}. Here we apply them to the Random Matrix Model and detail the technical aspects of the framework. 

In general since $\mu^2=\mu_{LY}^2$ is the closest singularity to the origin, the radius of convergence of the Taylor expansion in Eq. \ref{eq:eos_Taylor} is $|\mu_{LY}^2|$. However the coefficients $c_n(T)$ contain much more information than just the radius of convergence. The Darboux theorem \cite{darboux,Fisher:1974series} states that the behavior of the coefficients $c_n(T)$ at large order $n$ is directly related to the behavior of the function in the vicinity of the nearest singularity. Specifically, 
if the Taylor expansion coefficients of a function $f(z)=\sum_{n=0}^\infty b_n z^n$ near the origin have leading large-order growth as $n\to\infty$:
\bea
b_n\sim \frac{1}{z_0^n}
\left[
\binom{n+g-1}{n} 
\phi(z_0)- 
\binom{n+g-2}{n}
z_0\, 
\phi^\prime(z_0)+
\binom{n+g-3}{n}
\frac{z_0^2}{2!}\, \phi^{\prime\prime}(z_0)-\dots \right]
\label{eq:darboux2}
\ea
then the leading singularity is located at $z_0$, and in the vicinity of $z_0$ the function behaves as 
\begin{eqnarray}
f(z)\sim  \phi(z)\, \left(1-\frac{z}{z_0}\right)^{-g}+\psi(z) \qquad, \quad z\to z_0 
\label{eq:darboux1}
\end{eqnarray}
where $\phi(z)$ and $\psi(z)$ are analytic near $z_0$. This means that from a detailed study of the expansion coefficients $c_n(T)$, derived from the expansion about $\mu=0$, we can learn about the expansion of the function near the critical point. The leading term in \eqref{eq:darboux2} tells us the location of the singularity, as well as its exponent $g$ and its strength $\phi(z_0)$. The further subleading terms contain information about the expansion of $\phi(z)$ in the vicinity of the critical point. The practical question is: what is the most efficient way to extract as much of this information as possible, from a limited number of $c_n(T)$ coefficients? Optimized strategies for this problem have been developed recently in \cite{Costin:2020pcj,Costin:2021bay}.

\subsection{The \pade{} Approximant and the Radius of Convergence}

\begin{figure}[h]
\center
\includegraphics[scale=0.55]{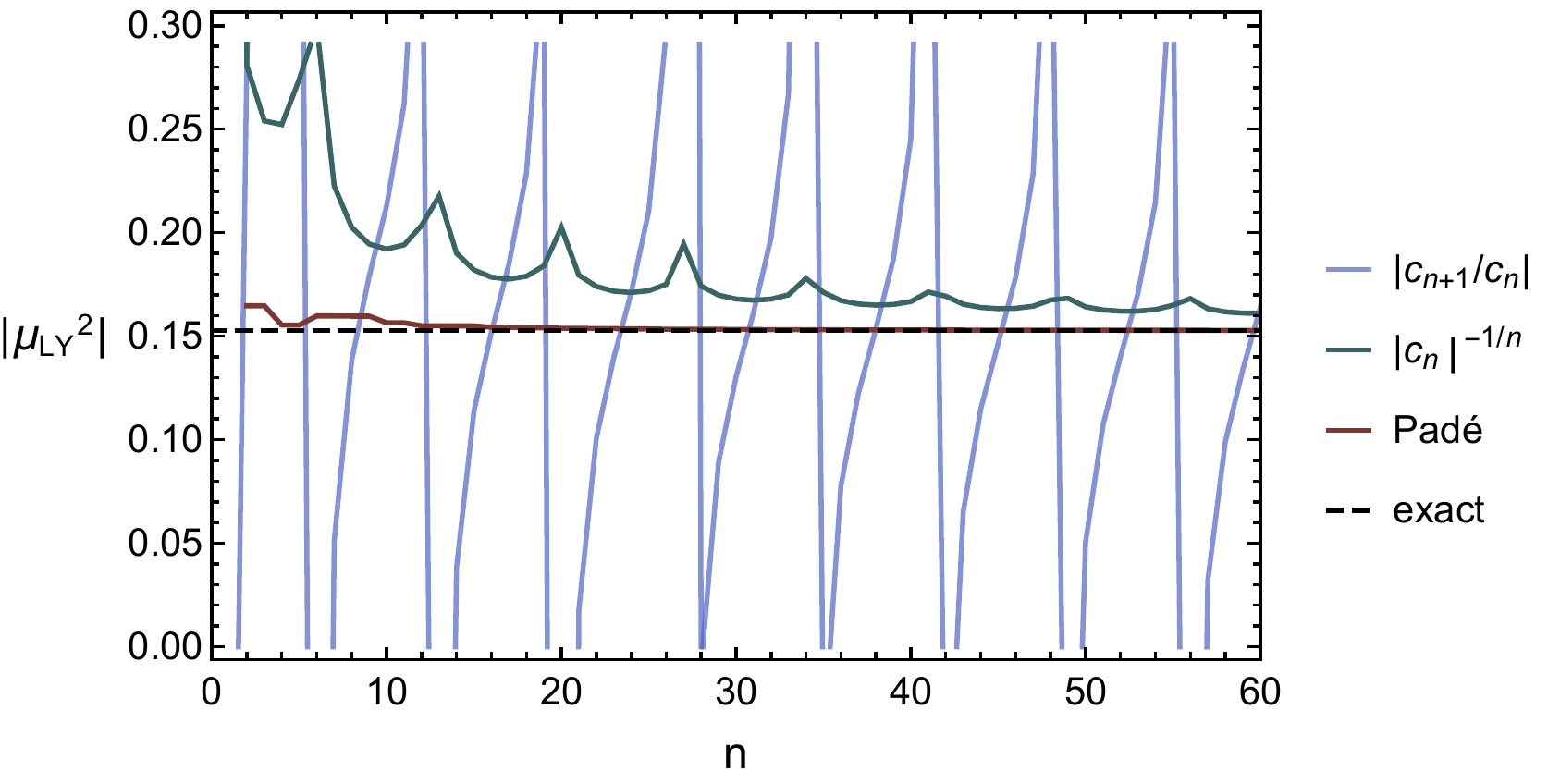}
\caption{The distance from the LY singularity to the origin, $|\mu_{LY}^2|$, extracted from a fixed number $n$ of coefficients $c_n(T)$ of the Taylor series expansion of the equation of state Eq. \eqref{eq:eos_Taylor}, using different methods. Here $m_q=0.1$ and $T=0.9$. The dashed horizontal line is the exact value at this temperature. The simple ratio test $|c_{n+1}(T)/c_n(T)|$ shows wild oscillations about the exact value, while the root test $|c_n(T)|^{1/n}$ is better behaved but converges slowly. Pad\'e methods are much more accurate with many fewer terms.
 }
\label{fig:muLYfromseries}
\end{figure}

Away from $T_c$, the Lee-Yang singularities occur as a complex conjugate pair. Since they are equidistant from the origin they both contribute, with equal weight but with phase of opposite sign. Hence the large-order growth of the expansion coefficients $c_n(T)$ exhibits oscillatory behavior at large order, due to interference between the complex conjugate pair of LY singularities. 
\bea
c_{n}(T)\sim  |\chi(\mu_{LY}(T))|
\binom{n-\sigma_{LY}-1}{n}
\frac{\cos\left(n\, \theta_{LY}(T) +\delta_{LY}(T)\right)}{|\mu_{LY}^2(T)|^n}
\quad, \quad n\to\infty
\label{eq:eos_large_order}
\ea
where $\theta_{LY}=\arg \mu_{LY}^2$, and $\delta_{LY}$ is the phase offset. The parameter $\sigma_{LY}$ determines the nature of the singularity.
This oscillatory behavior makes it numerically challenging to extract the relevant physical quantities (such as the location of $\mu_{LY}$, and the critical exponent, $\sigma_{LY}$) directly from ratio tests or root tests of the coefficients $c_n(T)$. See Figure \ref{fig:muLYfromseries}, in which we show the large order behavior of the ratio tests and the root tests of the coefficients, and compare with the exact radius of convergence for a fixed value temperature, $T\approx 1.34 T_c$. The interference effect is highly pronounced in the ratios which makes the extraction of $|\mu_{LY}^2|$ difficult. 
The root test  provides better convergence as the oscillating envelope is damped. However one still needs a large number of terms until this damping occurs. An alternative and significantly more efficient way to extract the singular behavior of the equation of state is to use Pad\'e approximants. Figure \ref{fig:muLYfromseries} shows significantly faster convergence for extracting the radius of convergence.

 Pad\'e methods provide an excellent, and simple, low-resolution probe of the singularity structure, and this can then be further refined by combining with conformal and uniformizing maps, as described below.
The idea is to approximate the original function by a rational function \cite{baker,bender},
\bea
{\rm P}[\chi(T,\mu)]=\frac{k_0(T)+k_1(T)\mu^2+\dots + k_{N/2}(T) \mu^{N} }{l_0(T)+l_1(T)\mu^2+\dots + l_{N/2}(T) \mu^{N} }\,.
\label{eq:eos_pade}
\ea
where we consider $N$ to be even. Here the coefficients $k_i$ and $l_i$ are determined by expanding Eq. \eqref{eq:eos_pade} in $\mu^2$ and matching with the original expansion, Eq. \eqref{eq:eos_Taylor}. This procedure is completely algorithmic, and is a built-in operation in symbolic software such as Mathematica and Maple.
It is worth noting that the label $T$ in Eq. \eqref{eq:eos_pade} should be viewed as an index rather than an argument of a smooth function, because the Pad\'e polynomials do not necessarily change smoothly when the coefficients $c_n(T)$ change smoothly. In general the order of the polynomials in the denominator and numerator could be different as long as there are $N+1$ independent coefficients. Here we consider the diagonal case where they have the same order unless specified otherwise, but it is important to note that numerical stability can be probed by varying the degrees of the Pad\'e polynomials. 
\begin{figure}[h]
\center
\includegraphics[scale=0.54]{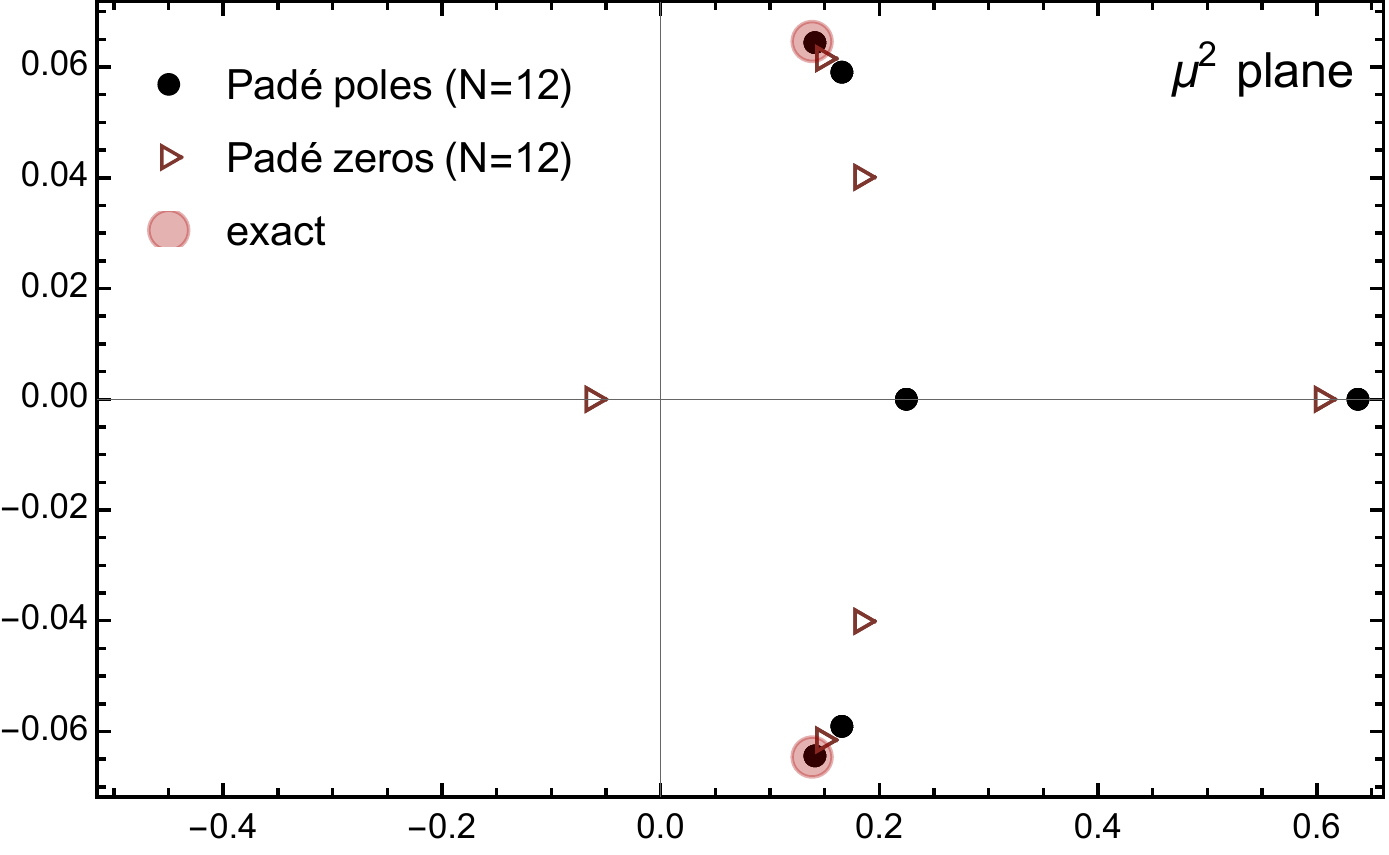}
\caption{The poles and zeros of Pad\'e approximants with 12 terms for $m_q=0.1$ and $T=0.9\approx 1.34 T_c$. Note the different scales on the two axes: the argument $\theta_{LY}$ of $\mu^2_{LY}$ is quite small.  In representing a complex conjugate pair of branch points, Pad\'e constructs a curved arc of poles and zeros joining the two branch points, together with another line of poles and zeros emanating from the first arc and extending to infinity along the real axis. This latter line is an artifact of Pad\'e, not a feature of the underlying function \cite{Stahl,aptekarev,Costin:2021bay}.}
\label{fig:pade-sing}
\end{figure}

\begin{figure}[h]
\center
\includegraphics[scale=0.55]{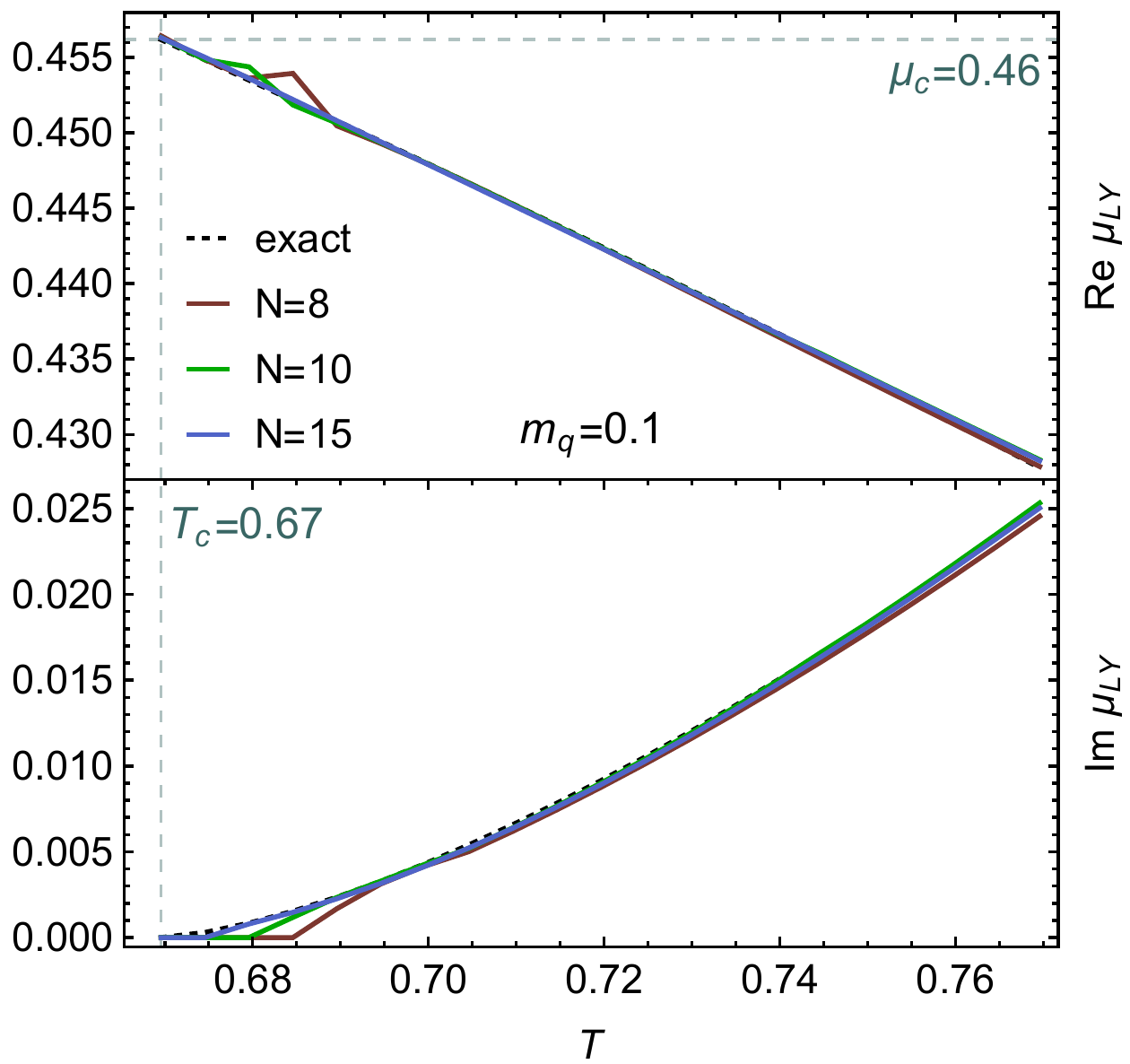}$\qquad\quad$
\includegraphics[scale=0.55]{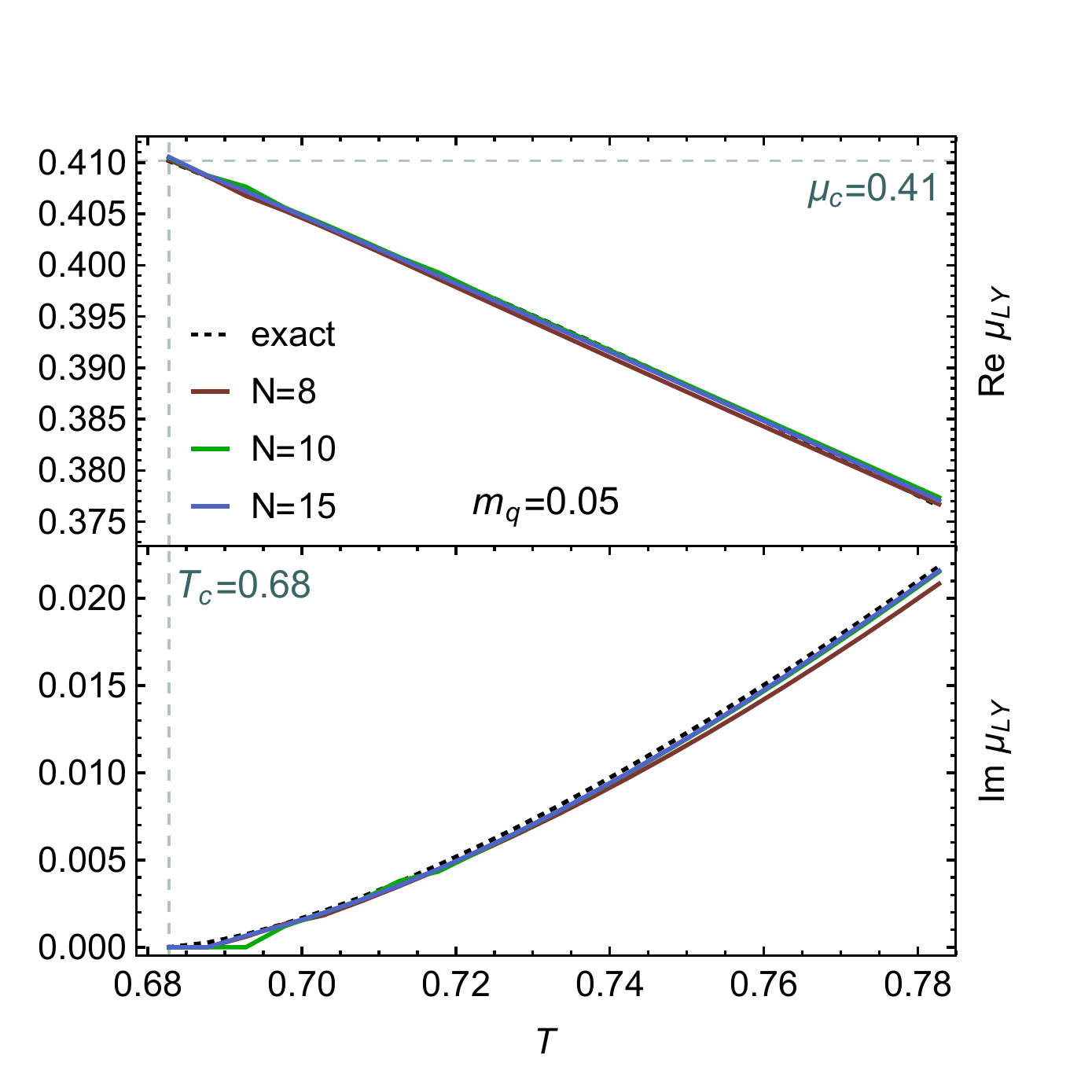}
\\
\includegraphics[scale=0.55]{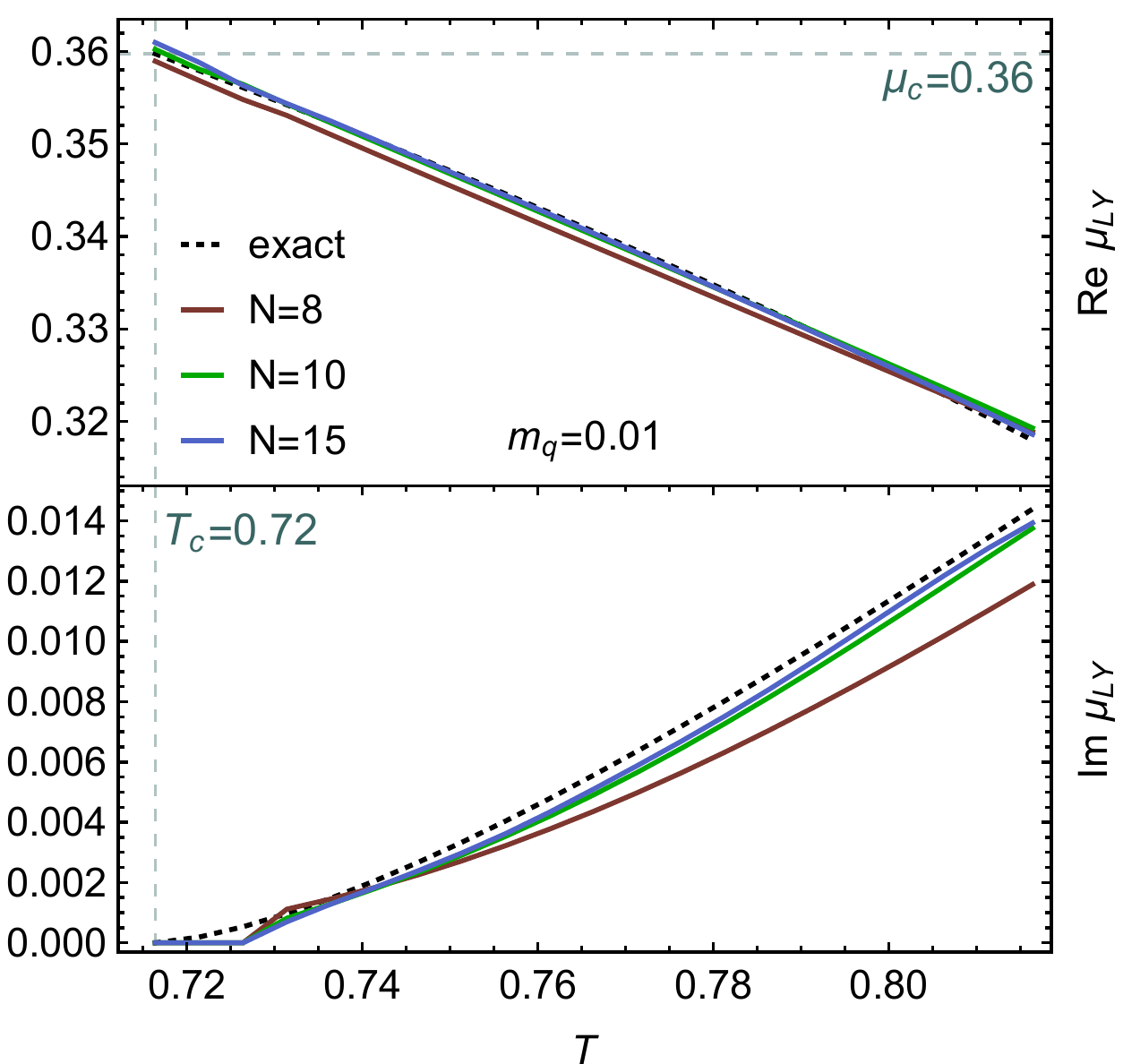}
$\qquad$
\includegraphics[scale=0.55]{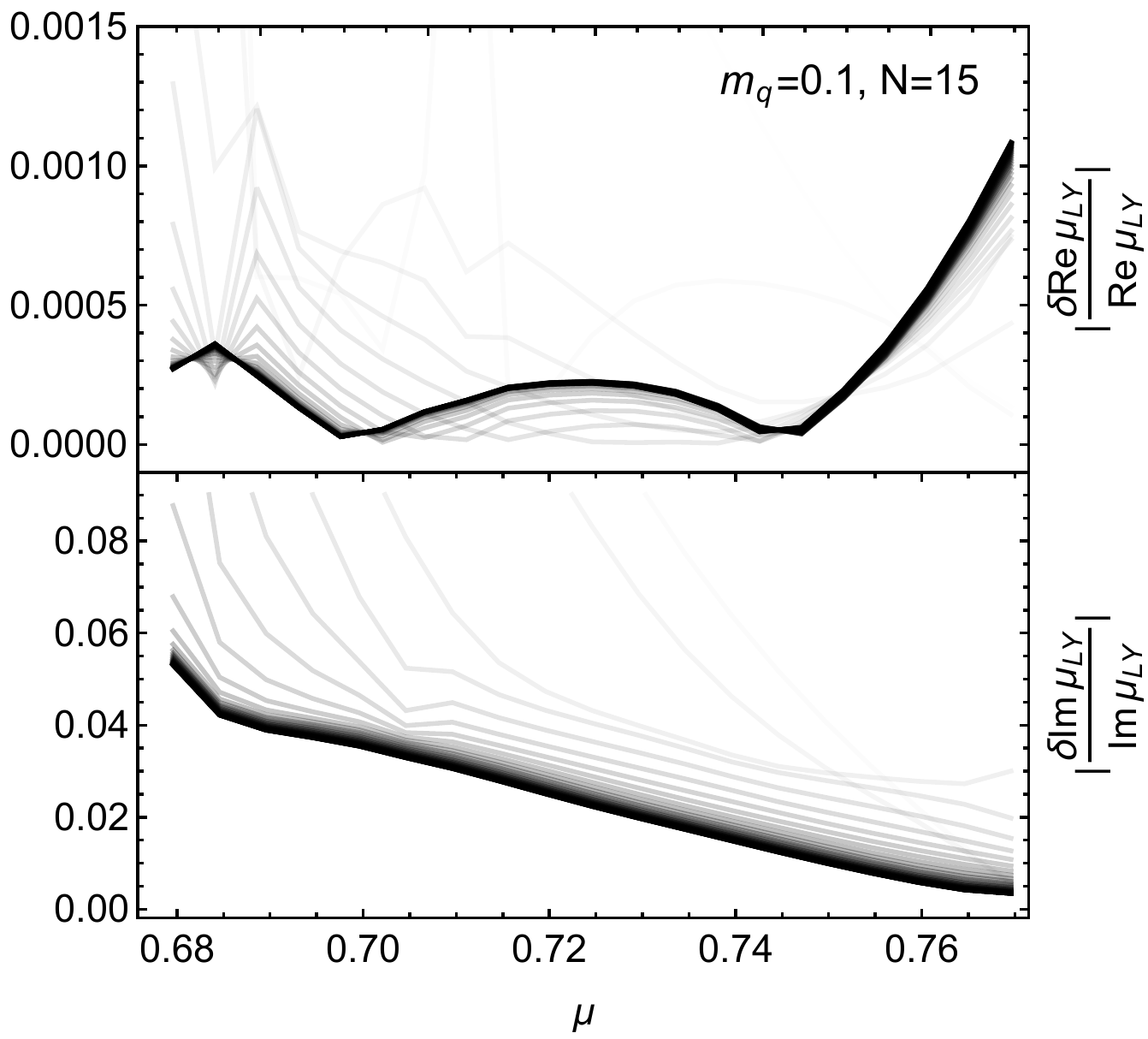}
\caption{The Lee Yang trajectory \eqref{eq:lytraj} extracted from conformal Pad\'e with different numbers of terms ($N=8, 10, 15$), and with quark masses ($m_q=0.1, 0.05, 0.01$), compared with the exact result. Bottom-right: The convergence of the iterative procedure (explained in text) with 30 iteration steps represented by the lines becoming darker.}
\label{fig:ly-traj-cpade}
\end{figure}

In contrast with the truncated Taylor expansion \eqref{eq:eos_Taylor}, the corresponding Pad\'e approximant has poles.  Along with the zeros they approximate the location of the leading singularity. For a function with branch points, Pad\'e represents the branch points as the accumulation points (in the $N\to\infty$ limit) of arcs of interlacing poles and zeros, whose shape is determined by minimizing an effective electrostatic capacitance \cite{Stahl,aptekarev,Costin:2020pcj,Costin:2021bay}. 
We see in Figure \ref{fig:pade-sing} that even with relatively few coefficients ($N=12$), the branch point locations in the complex plane can be estimated very accurately. Below we describe an iterative method to refine this initial estimate to even higher precision. See Fig. \ref{fig:ly-traj-cpade}. Figure \ref{fig:muLYfromseries} shows the value of $|\mu_{LY}^2|$ extracted from Pad\'e resummation, plotted as a function of the number of input  coefficients. We chose the pole that is closest to origin and in order to eliminate some of the unphysical singularities filtered out those with residues smaller than $10^{-6}$. The results are stable under variations of the cutoff and the Pad\'e orders. In Figure \ref{fig:muLYfromseries} we see that the convergence to the exact result is much faster and more accurate than either the ratio or root tests. Furthermore, Figure \ref{fig:pade-sing} shows that the Pad\'e approximant gives direct access to the full complex form of $\mu_{LY}^2$, not just the magnitude $|\mu_{LY}^2|$. It is also important to note that this result is obtained by using the \textit{same input} as the truncated Taylor expansion, namely just the series coefficients, without any additional input.

In addition to providing a significantly better way of locating its singularities, Pad\'e approximants can also be used to improve the accuracy of the approximation to the equation of state and the susceptibilities. In Fig. \ref{fig:pade-chis} we show the result of the Pad\'e approximant for the susceptibility, as well as for the higher order susceptibilities 
\bea
\chi_n(T,\mu)=\frac{\del^n p(T,\mu)}{\del\mu^n}\,
\ea
These observables, especially $\chi_3$ and $\chi_4$, play a crucial role in the search for the critical point as their magnitude grows in the vicinity of the critical point. Their counterparts in QCD are related to the skewness and kurtosis of the net baryon number distribution \cite{best}. Furthermore, their shape and quantitative features depend non-trivially on the mapping parameters given in Eq. \eqref{eq:params} \cite{Pradeep:2019ccv}. Therefore it is vitally important to obtain an accurate approximation to the equation of state in order to resolve this structure. As seen in Fig. \ref{fig:pade-chis}, the Pad\'e approximant slightly outperforms the truncated Taylor series, which diverges at $\mu^2=|\mu_{LY}^2|$ as expected. However, Pad\'e also diverges for $\mu^2 \gtrsim |\mu_{LY}^2|$. This is because it hits an unphysical pole along the real axis.

 The existence of these unphysical poles, as highlighted in Fig. \ref{fig:pade-sing}, is an inherent shortcoming of Pad\'e when there is a complex conjugate pair of branch points. For such a configuration, in minimizing the effective capacitance, Pad\'e places spurious unphysical poles and zeros along the positive real axis, extending out to infinity and becoming dense as $N\to\infty$ \cite{Stahl,aptekarev,Costin:2020pcj,Costin:2021bay}.\footnote{These spurious poles and zeros are distinct from other spurious poles and zeros due to numerical instabilities from lack of precision.} The effect of these spurious  poles and zeros can be seen in Fig. \ref{fig:pade-chis}, where they are responsible for the large unphysical oscillations of the Pad\'e-reconstructed susceptibilities beyond the radius of convergence. This unphysical effect can be overcome by using conformal or uniformizing maps in conjunction with Pad\'e, as described below, thereby significantly extending the range of applicability of the resummation.

\subsection{The Conformal \pade{} Method}

Up to this point we have not used any additional information other than the expansion coefficients. We could further improve the accuracy of our approximation by providing more information. For example, we know that for $T>T_c$, the equation of state has two complex conjugate branch points, $|\mu_{LY}^2|e^{\pm i\theta}$ \cite{Stephanov:2006dn,An:2017brc} as well as a critical exponent, even beyond the mean field limit. We can incorporate this information before making the Pad\'e approximation. 

\begin{figure}[h]
\center
\includegraphics[scale=0.7]{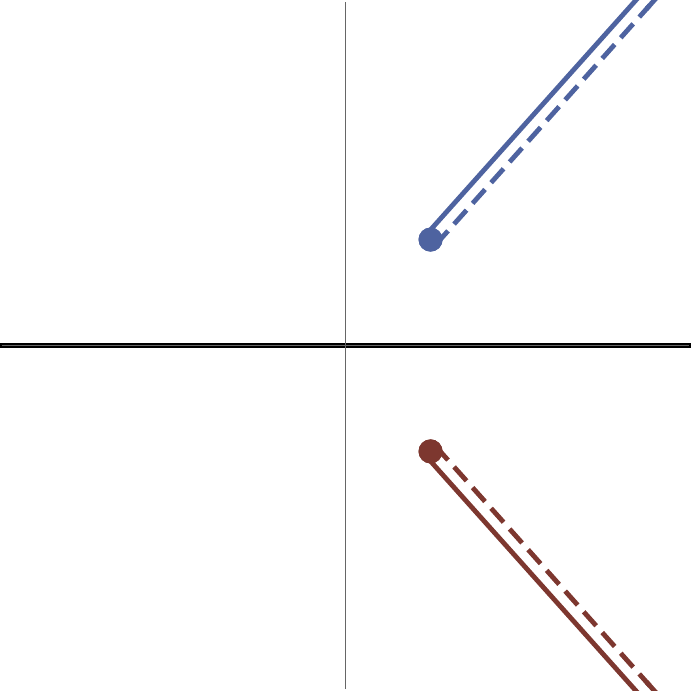}
$\qquad\qquad\qquad$
\includegraphics[scale=0.7]{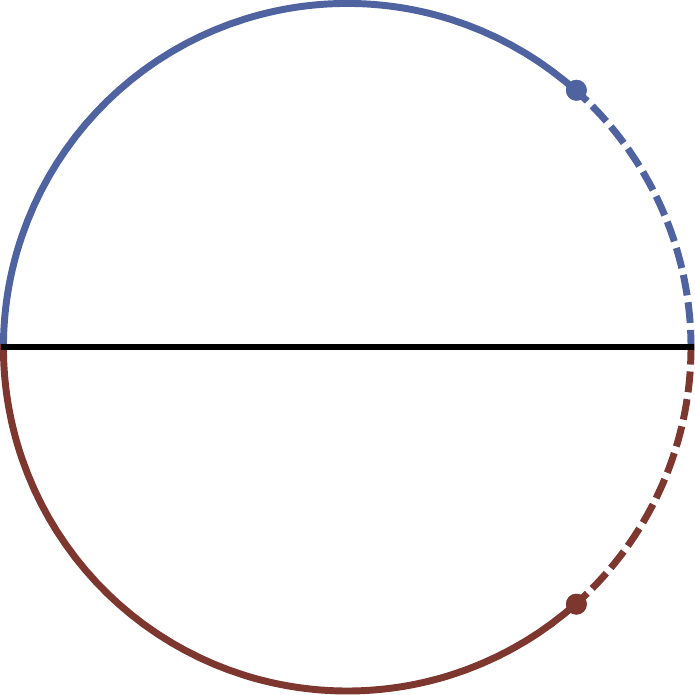}
\caption{The singularity structure in the $\mu^2$ plane for $T>T_c$ where the dots denote the LY singularities and the lines denote symmetrically chosen branch cuts (left) and its image on the unit disk under the conformal map Eq. \eqref{eq:conf_map} (right).}
\label{fig:conf_map}
\end{figure}

The first step is to conformally map the $\mu^2$ plane into the unit disk ($|\zeta|<1$) by identifying $\mu^2$ with $\phi(\zeta)$.  Conformal maps have well known applications in summation methods \cite{ZinnJustin:2002ru,Caliceti:2007ra,caprini,Costin:2021bay}. Here we need a conformal map for a complex conjugate pair of singularities, as shown in Figure \ref{fig:conf_map} [left]. 
We use the conformal map, $\phi(\zeta)$, where 
\bea
\phi(\zeta)=4 |\mu_{LY}^2| \left(\frac{\theta}{\pi}\right)^{\theta/\pi}
\left(\frac{1-\theta}{\pi}\right)^{1-\theta/\pi}
\frac{\zeta}{(1 + \zeta)^2}\left(\frac{1 + \zeta}{1 - \zeta}\right)^{2 \theta/\pi}
\label{eq:conf_map}
\ea
This maps the $\mu^2$ plane into the unit disk as shown in Fig. \ref{fig:conf_map}.\footnote{ See \cite{nehari,kober}. This particular map appears in a wide range of physical applications \cite{Rossi:2018urw,Serone:2019szm,bertand}.} The branch points (the LY singularities) and the associated cuts (red, blue dots and lines) are mapped onto the boundary of the unit disk. Notice that each side of each cut (depicted as a dashed or solid line) is mapped to a different segment of the unit circle, meeting at the associated branch point and at $\zeta=\pm 1$, corresponding to the point at infinity.

The next step is to re-expand the susceptibility $\chi(T,\phi(\zeta))$ as a series in $\zeta$ instead of as a series in $\mu^2$, truncating at the same order (this procedure is optimal \cite{Costin:2020pcj}): 
\bea
\chi(T,\phi(\zeta))\approx\sum_{n=0}^{N} \tilde c_n(T)\zeta^n
\label{eq:conf_taylor}
\ea
followed by a Pad\'e approximant (now in terms of $\zeta$) of this re-expansion. 
Let us denote the poles and zeros of this modified Pad\'e approximant as $\zeta_i$ for $i=1,\dots, N$. Using the conformal map \eqref{eq:conf_map} we map them back to the $\mu^2$ plane:
\bea
\mu_i^2=\phi(\zeta_i)\,.
\label{eq:conf_pade_sing}
\ea
Similar to Pad\'e singularities, the $\mu_i^2$ values accumulate towards the physical singularities, $\mu_{LY}^2$. 
\begin{figure}[h]
\center
\includegraphics[scale=0.47]{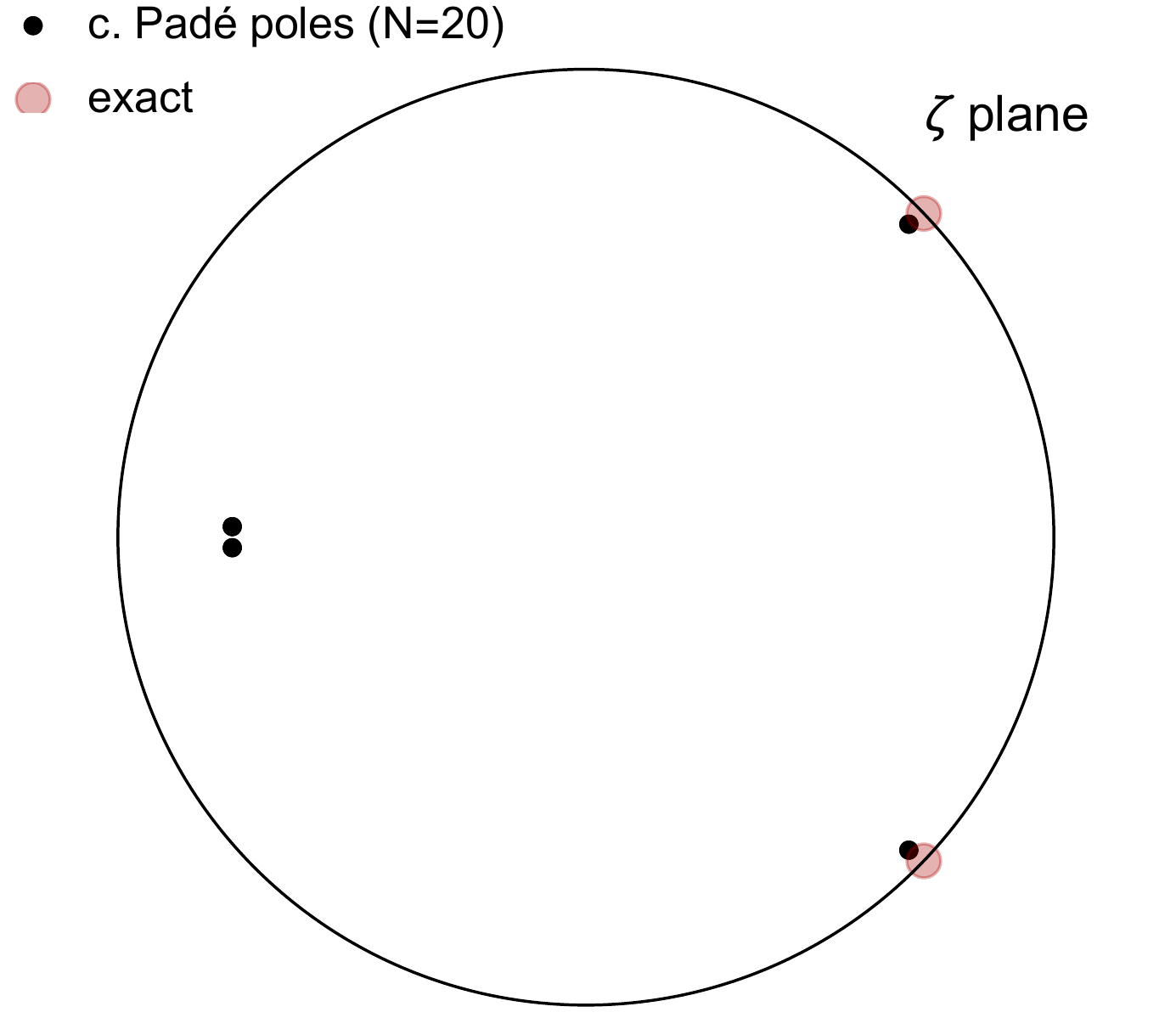}
$\qquad\qquad$
\includegraphics[scale=0.55]{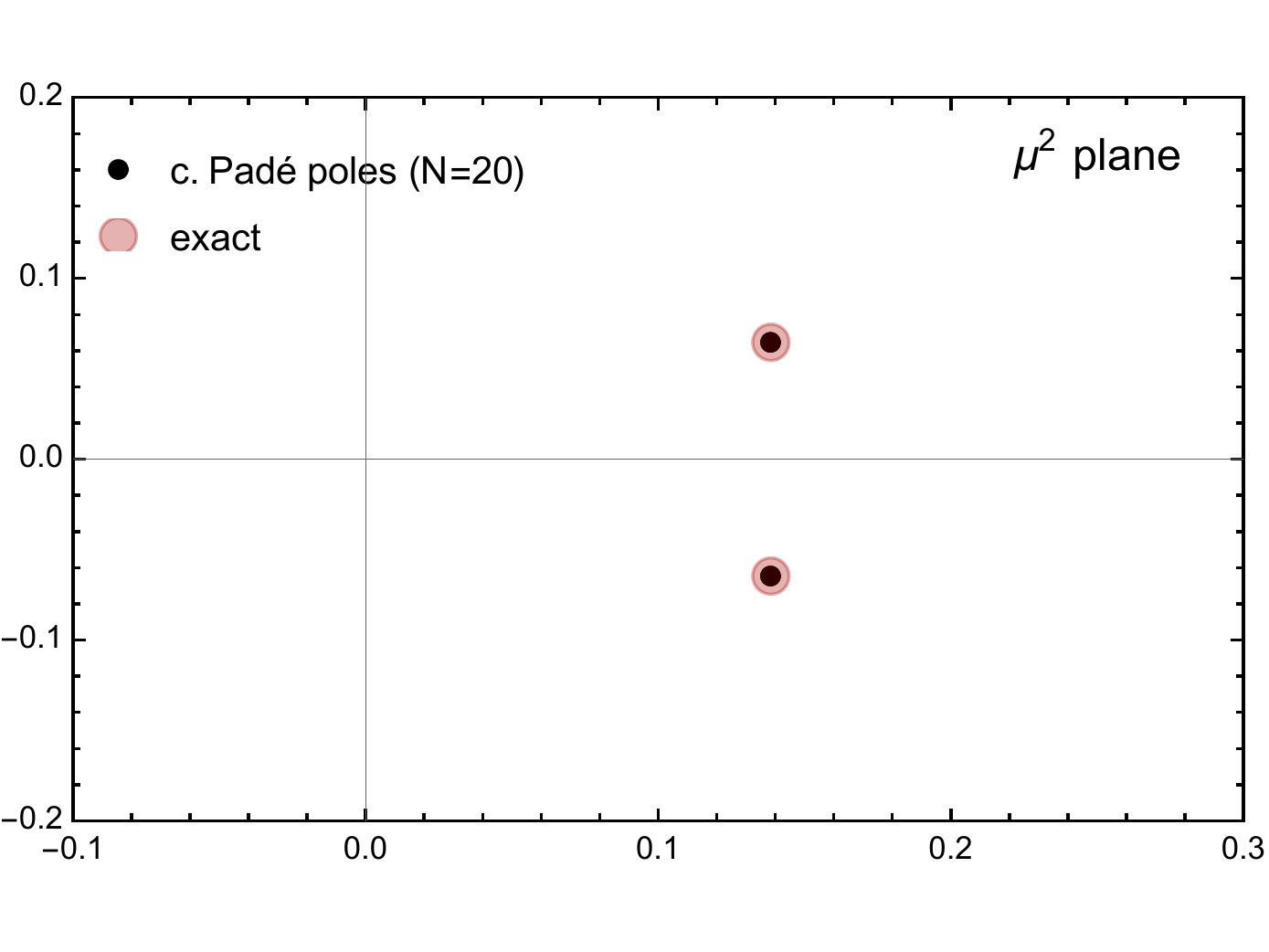}
\caption{The poles of conformal Pad\'e in the unit disk (left), and mapped back to the $\mu^2$ plane (right). The conformal Pad\'e method locates the physical LY singularities with high precision. These plots are made for $m_q=0.1$ and $T=0.9\approx 1.34 T_c$.}
\label{fig:cpade-sing}
\end{figure}

In Fig. \ref{fig:cpade-sing} we show the singularities in the unit disk and in the $\mu^2$ plane mapped via Eq. \eqref{eq:conf_map}. Notice that compared with the exact result, the accuracy both in the unit disk and in the $\mu^2$ plane is remarkable. 
This increase in precision can be quantified as a function of the number of terms \cite{Costin:2020hwg}. Note that the conformal map \eqref{eq:conf_map} only requires knowledge of the {\it location} of the singularities, not of their exponent or strength. Therefore we can iterate this procedure to obtain even higher precision, which is particularly useful if only few coefficients are known (see the next section). Also notice that as opposed to Pad\'e, there are no unphysical singularities along the real axis. This is due to the asymptotic form of the Pad\'e polynomials, explained in \cite{Costin:2020hwg}.

\begin{figure}[h]
\center
\includegraphics[scale=0.5]{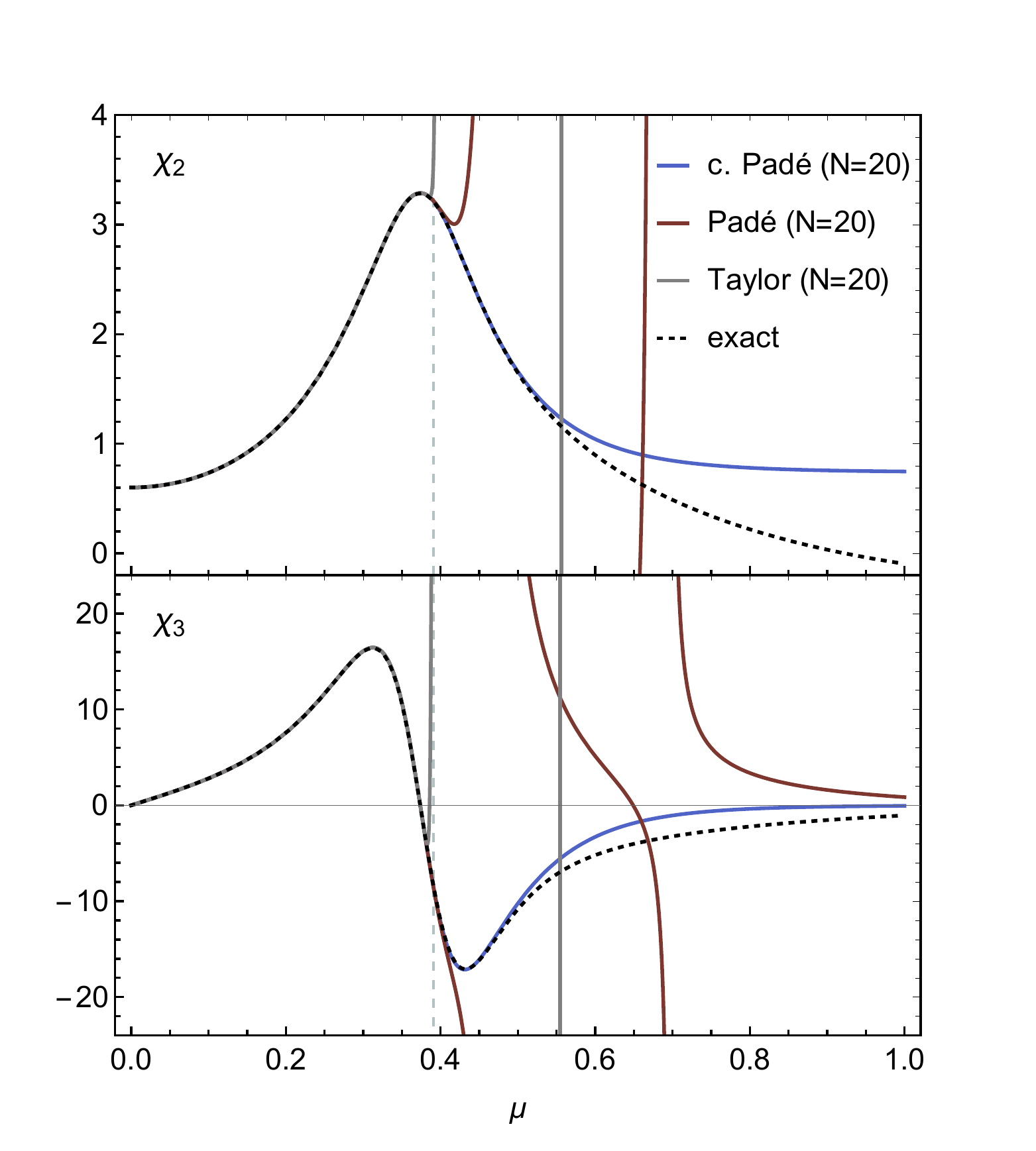}
\includegraphics[scale=0.5]{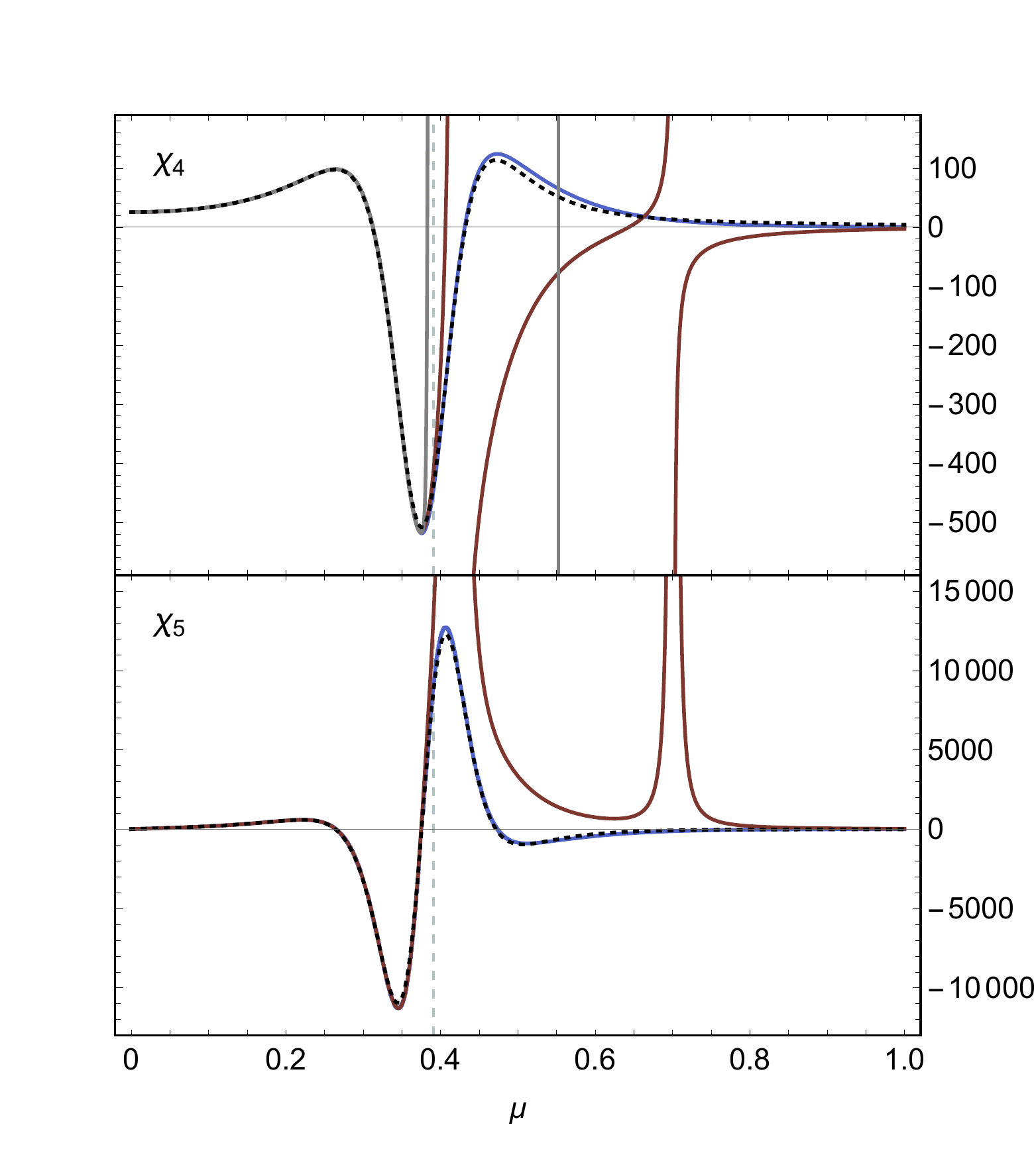}
\caption{The susceptibilities $\chi_2$,$\chi_3$,$\chi_4$,and $\chi_5$ for $m_q=0.1$ and $T=0.9\approx 1.34 T_c$ obtained from directly summing the truncated series expansion [gray], Pad\'e [red] and conformal Pad\'e [blue] resummations, compared with the exact result [black dashed]. The vertical dashed gray line denotes the radius of convergence $\mu=|\mu_{LY}|\approx 0.39$. Note that only the conformal Pad\'e reconstruction is able to resolve the higher-order structure of the susceptibilities.}
\label{fig:pade-chis}
\end{figure}

In addition to the location of $\mu_{LY}^2$, the susceptibility can be reconstructed from the conformal Pad\'e expression mapped back to the $\mu^2$ plane:
\bea
\chi(T,\mu)\approx {\rm P}[\chi(T,\phi(\zeta))]|_{\zeta=\phi^{-1}(\mu^2)}
\label{eq:conf_pade}
\ea
Note that apart from certain special rational values of $\theta$, there is no analytic expression for the inverse function $\phi^{-1}(\mu^2)$, but it is straightforward to implement  this inversion numerically. This procedure provides a significantly superior approximation to the susceptibility compared to ordinary Pad\'e. This can be seen in Fig. \ref{fig:pade-chis}, where we compare the conformal Pad\'e results with other methods:  Pad\'e and the truncated Taylor expansion. The most dramatic improvement is in the range of validity of the extrapolation. The lack of unphysical poles along the real axis allows conformal Pad\'e to provide a much better approximation to the susceptibility much further than the radius of convergence of the original Taylor series, $|\mu_{LY}^2|$. In particular, the qualitative features of the higher order susceptibilities, such as the ``peak-dip" structure of $\chi_3$, and the ``peak-dip-peak" structure of $\chi_4$, are successfully reproduced even with a relatively small number of coefficients. These features cannot be seen in the truncated Taylor series, or in its Pad\'e approximant. We stress that exactly the same input information (the coefficients $c_n(T)$) was used for these three approximations: this input data was simply processed differently, with the conformal Pad\'e procedure being clearly superior.

\subsection{Reconstruction of the Lee-Yang trajectory}
We repeated the procedure explained above for a range of temperatures to reconstruct the Lee Yang trajectory in Eq. \eqref{eq:lytraj}, using the  conformal Pad\'e resummation. As mentioned earlier, the conformal map Eq. \eqref{eq:conf_map} contains $\mu_{LY}$, the quantity we wish to compute from it. We therefore follow an iterative procedure which goes as follows:
\begin{enumerate}
\item Estimate a preliminary $\mu_{LY}^2$ from ordinary Pad\'e. This step does not involve the conformal map. 
\item Plug this value in the conformal map, Eq. \eqref{eq:conf_map}, and perform conformal 
Pad\'e.
\item Update the value of $\mu_{LY}^2$ extracted from conformal Pad\'e.
\item Go to step 2 and repeat.
\end{enumerate}

In steps 1 and 3, among the poles and zeros of Pad\'e and conformal Pad\'e we select the one that best approximates \muLYsq{}. 
This is achieved by filtering the poles and zeros to be stable under variation of the orders of the Pad\'e polynomials.
Then, the one with the largest imaginary part is selected from the filtered poles and zeros. If all poles and zeros are on the real axis, the one closest to the origin is selected. 

The refined estimation $\mu^2_{EST}(T)$ so obtained may converge to outlying values for certain temperatures. To mitigate this uncertainty, a second-stage local optimization is performed by first fitting $\mu^2_{EST}(T)$ to the formula in Eq. \eqref{eq:lytraj} (excluding obvious outliers) and inputting the fitted values $\mu^2_{FIT}(T)$ to the conformal map. A refined value for $\mu^2_{EST}(T)$ is then obtained by selecting the re-expanded Pade pole or zero closest to $\mu^2_{FIT}(T)$ in the $\mu^2$ plane, and the whole process is repeated until convergence.

\begin{table}[h]
\center
\begin{tabular}{ |c | c| c| c| } 
\hline
$m_q:$ & 0.1 &0.05 & 0.01
\\
\hline
$T_c,\mu_c$ (exact):$\quad$ &0.670, 0.456 & 0.683, 0.410  & 0.716, 0.360
\\
\hline
$T_c,\mu_c$ ($N=15$): &0.670, 0.456 & 0.684, 0.410  & 0.719, 0.360
\\
\hline
$T_c,\mu_c$ ($N=8$):\,\,\, &0.674, 0.455 & 0.683, 0.410  & 0.717, 0.359 \\
\hline
\hline
$K_1,K_2$ (exact):$\quad$ &-0.267, 0.844 & -0.308, 0.740  &-0.362, 0.539
\\
\hline
$K_1,K_2$ ($N=15$): &-0.278, 0.831 & -0.325, 0.719  & -0.397, 0.526 
\\
\hline
$K_1,K_2$ ($N=8$):\,\,\, &-0.282, 0.979 & -0.334, 0.699  & -0.400, 0.421
\\
\hline
\end{tabular}
\caption{The location of the critical point, $(T_c, \mu_c)$, and the Ising mapping parameters, $(K_1, K_2)$ of Eq. \eqref{eq:params}, obtained from the conformal Pad\'e reconstruction of the LY trajectory, using $N=15$ and $N=8$ input coefficients in the initial expansion \eqref{eq:lytraj}. For comparison we also list the exact values.  }.
\label{table}
\end{table}

 We used 30 steps for both stages of the iterative procedure. In Fig. \ref{fig:ly-traj-cpade} (top and bottom-left) we show the LY trajectory constructed by this procedure, for different values of $m_q$. In Fig. \ref{fig:ly-traj-cpade} (bottom-right) we show the convergence of the second-stage iterative procedure for $m_q=0.1$ and $N=15$. For visualization purposes the opacity of the curves is changed with the iteration step, becoming darker as the iterative procedure progresses. As seen from the Figure, we obtain the real part of the Lee-Yang singularity roughly with $0.02\%$ accuracy. As expected, it is more difficult to resolve the imaginary part, as it vanishes at the critical point. However the accuracy is still at the percent level.

Note that even with just 8 terms in the initial expansion, the agreement with the exact result is quite good. The biggest challenge arises in the region very close $T=T_c$ where $\im \mu_{LY}$ approaches zero and resolving the small imaginary part becomes more difficult numerically. At the same time, as seen in Fig. \ref{fig:ly-traj-cpade}, it is still possible to capture the $(T-T_c)^{3/2}$ behavior for $T\gtrsim T_c$ even for $N=8$. This is in contrast to ordinary Pad\'e which provides a poor resolution of $\im \mu_{LY}$ near $T_c$ \cite{PhysRevLett.127.171603}. Finally by fitting the curves in Fig. \ref{fig:ly-traj-cpade} to the expected form of the trajectory, \eqref{eq:lytraj} $T_c$,$\mu_c$, as well as the coefficients $K_1$ and $K_2$. The results are given in Table \ref{table}. They are in good agreement with the
 exact results calculated directly from the mapping parameters $(h_\mu,h_T,r_\mu,r_T)$ obtained via the Ginzburg-Landau analysis explained in Sec. \ref{sec:rmm}.

\section{Uniformization and analytic continuation of the Ising equation of state}
\label{sec:uniform}

In the previous section we showed how to extract highly accurate physical information about the Lee-Yang singularities from a finite-order polynomial approximation to the expansion of the partition function (or susceptibility) in powers of the chemical potential. In this section we discuss an even more difficult problem: how to extrapolate information in the high temperature region, $T>T_c$, to the low temperature region, $T<T_c$. This requires analytic continuation from the first Riemann sheet ($T>T_c$), where the original expansion is generated, across a cut to the next Riemann sheet, where $T<T_c$. We demonstrate that this can be achieved even when starting with a finite-order expansion, using methods developed in \cite{Costin:2020pcj,Costin:2021bay}. This requires going beyond simple Pad\'e analysis and conformal maps, instead using uniformizing maps which encode more information about branch cut structures. In this section we illustrate these ideas on the mean field Ising model and, in particular, show that the equation of state for low temperatures, $T<T_c$, can be constructed from the high temperature expansion at $T>T_c$ via analytic continuation. We first show that an {\it exact} uniformization is possible for the mean field Ising model, and then we show that even with partial information a simple uniformizing map enables accurate analytic continuation between Riemann sheets.

The key idea behind uniformization is to map the entire multi-sheeted domain of the original function to the upper half plane (it is useful to then map to the unit disk) in a specially chosen uniformizing variable \cite{bateman,abikoff,Costin:2020pcj,Costin:2021bay}. The net result is that different sheets are mapped to different regions of the unit disk, whose boundaries are connected by modular transformations. If the exact uniformizing map is known then this procedure is optimal and explicit \cite{Costin:2020pcj,Costin:2021bay}. 
While it is rare in non-trivial physics problems to know the {\it exact} uniformizing map for the underlying Riemann surface, fortunately this uniformization procedure can also be implemented numerically using {\it approximate} information about the Riemann surface. For example, even an approximate uniformizing map, simply based on the locations of a few leading singularities, leads to dramatically higher precision on the first sheet, as well as the ability to cross approximately to other sheets \cite{Costin:2020pcj,Costin:2021bay}. This use of approximate information about leading singularities is analogous to well-known procedures combining conformal maps with Pad\'e analysis \cite{ZinnJustin:2002ru,Caliceti:2007ra,caprini}. However, there is an important difference: with a conformal map one is limited to a given sheet, as the original sheet is mapped inside the whole unit disk. If instead one uses a uniformizing map to map first to the upper-half-plane and then into the unit disk, the first sheet is mapped to a particular region of the unit disk, and second sheet to another (connected) region, and so on. Therefore, analytic continuation trajectories in the unit disk can pass smoothly between sheets. In such a situation the uniformizing map is generically much more accurate than a corresponding conformal map, and dramatically more accurate than a Pad\'e approximation. 

Since the Ising model system has certain universal features, such as the existence of a dominant pair of complex conjugate singularities, even the mean field analysis can be used to develop suitable approximate uniformizing maps which turn out to be significantly more accurate. We compare an exact uniformization of the mean field Ising system with an approximation based on a truncated initial expansion. The method is extremely simple to implement (see Section \ref{sec:optimal} below), and can in principle be adapted to more general problems, and beyond mean field.

\subsection{Exact uniformization of the Ising model equation of state}
\label{sec:exact}

Before discussing the analytic continuation through resummation of an approximate truncated expansion, let us briefly elaborate further on the analytic structure of the Ising equation of state. The naive solution $z=z(w)$ of the equation of state \eqref{eq:eos_zw} produces three different expressions for $z(w)$ involving complicated cube roots. This reflects the fact that the solution to the equation of state is defined on a three-sheeted Riemann surface, which is in turn a direct consequence of truncating the action \eqref{eq:action} at $\mathcal O(\phi^4)$.

However, as we show below, the mean field equation of state can also be solved in terms of a uniformizing variable, which makes the transition between sheets transparent.
\begin{figure}[h]
\center
\includegraphics[scale=0.6]{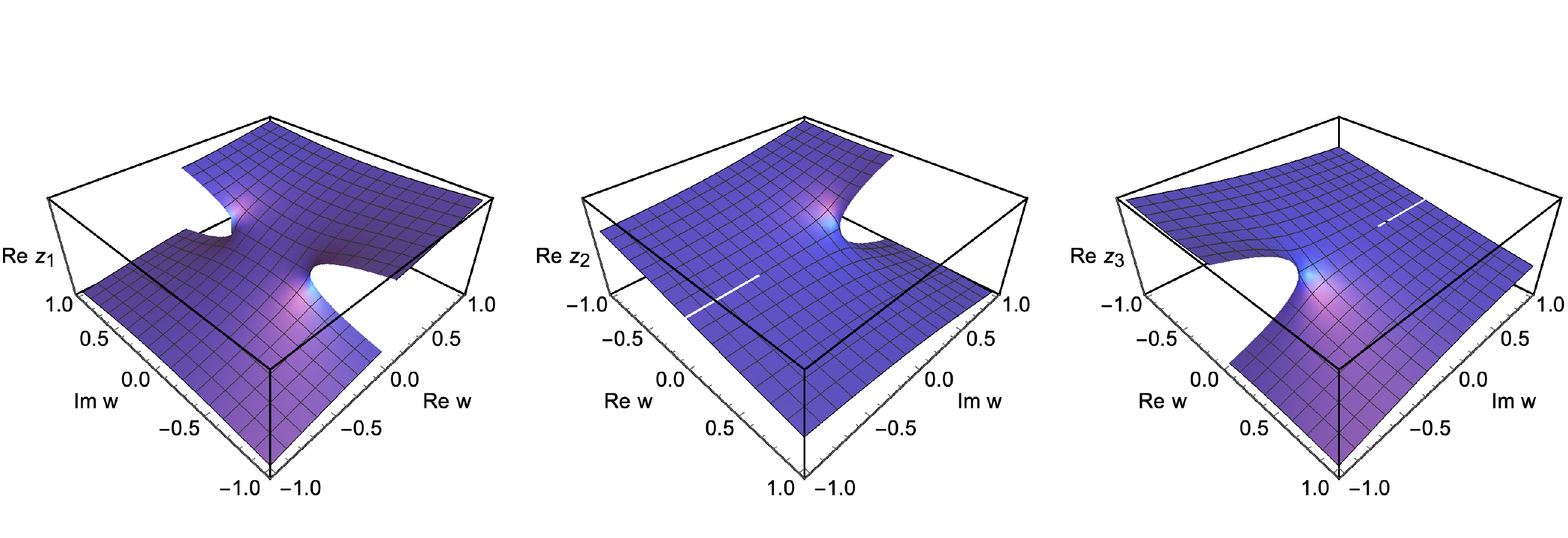}
\caption{The solution $\re z(w)$ of the scaled equation of state \eqref{eq:eos_zw} in the complex $w$ plane. From left to right: $\re z_1(w)$, $\re z_2(w)$ and $\re z_3(w)$. Similar plots can be generated for the imaginary part.}
\label{fig:eos3d}
\end{figure}

\begin{figure}[h]
\center
\includegraphics[scale=0.5]{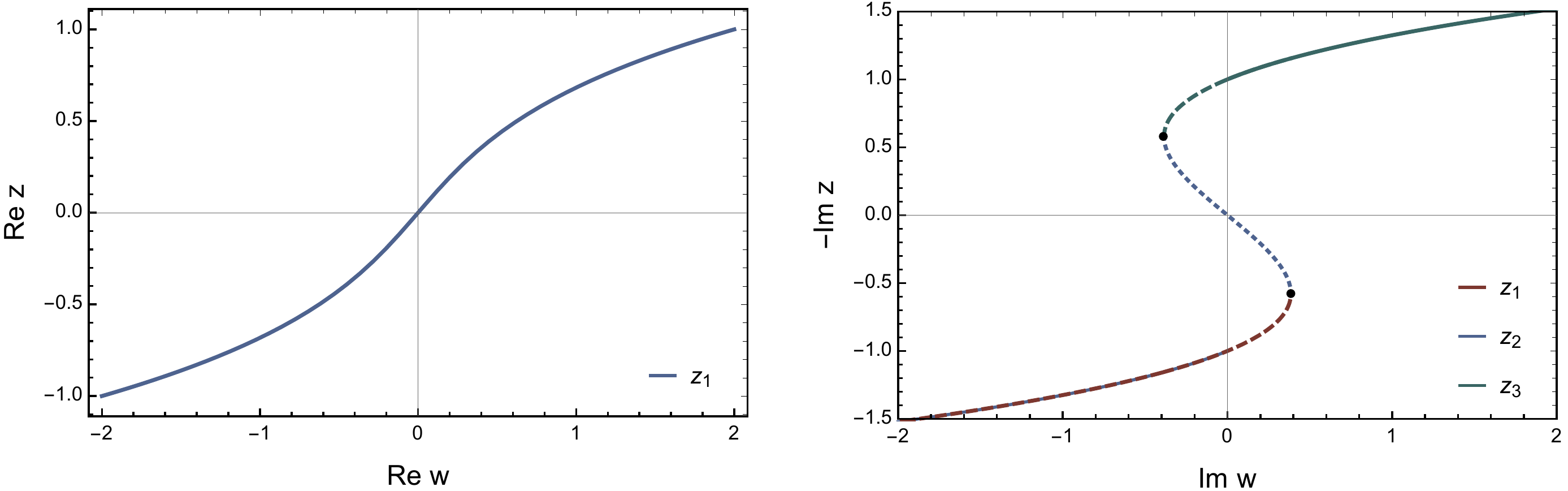}
\caption{The scaled equation of state $z(w)$ for $T>T_c$ (left) and $T<T_c$ (right). The black dots in the right-hand figure show the location of the Lee-Yang singularities $w=\pm\frac{2i}{3\sqrt{3}}$ which corresponds to the spinodal point. The dashed/dotted parts denote the meta-stable and unstable regions, respectively.}
\label{fig:eos_zw}
\end{figure}

To set notation, let us denote the three solutions of the scaled equation of state \eqref{eq:eos_zw} as $z_1(w),z_2(w)$ and $z_3(w)$, each defined over one of the sheets (see Fig. \ref{fig:eos3d}). Only two of the solutions are independent as $z_3(w)=-z_2(-w)$. We will refer to the sheets over which $z_1(w)$ and $z_2(w)$ are defined as the high $T$ and low $T$ sheets, respectively. The high $T$ sheet captures the equation of state for $r=(T-T_c)/T_c>0$ (see Fig. \ref{fig:eos_zw}, left). It has two branch cuts emanating from branch points at $w=\pm 2i/(3\sqrt{3})$, which correspond to the Lee-Yang edge singularities where $dw/dz=0$. The other two sheets over which $z_2(w)$ and $z_3(w)$ are defined capture the low temperature, $T<T_c$, behavior where $r<0$. They are related to each other by the reversal of the direction of the magnetic field, $h$, i.e. $w\rightarrow -w$. The low $T$ sheets each have a single branch point, at $z=\pm 2i/(3\sqrt{3})$, respectively. Furthermore one can move from the high $T$ sheet to the low $T$ sheet via the analytic continuation: $r\rightarrow e^{-i\pi} r$, which corresponds to $w\rightarrow e^{3i\pi/2} w$ and $z\rightarrow e^{i\pi/2}$. The equation of state, $M(h)$ for $T>T_c$ and $T<T_c$ is shown in Fig. \ref{fig:eos_zw}, in terms of the scaling variables $z$ and $w$. For $T>T_c$ the physical (i.e. real) values of magnetization and magnetic field correspond to $\re z_1$ and $\re w$, respectively. For $T<T_c$ and $h<0$, after analytic continuation we obtain $M\propto -\im z_2$. Similarly for $T<T_c$ and $h>0$ we have $M\propto -\im z_3$. 

\begin{figure}[h]
\center
\includegraphics[scale=0.6]{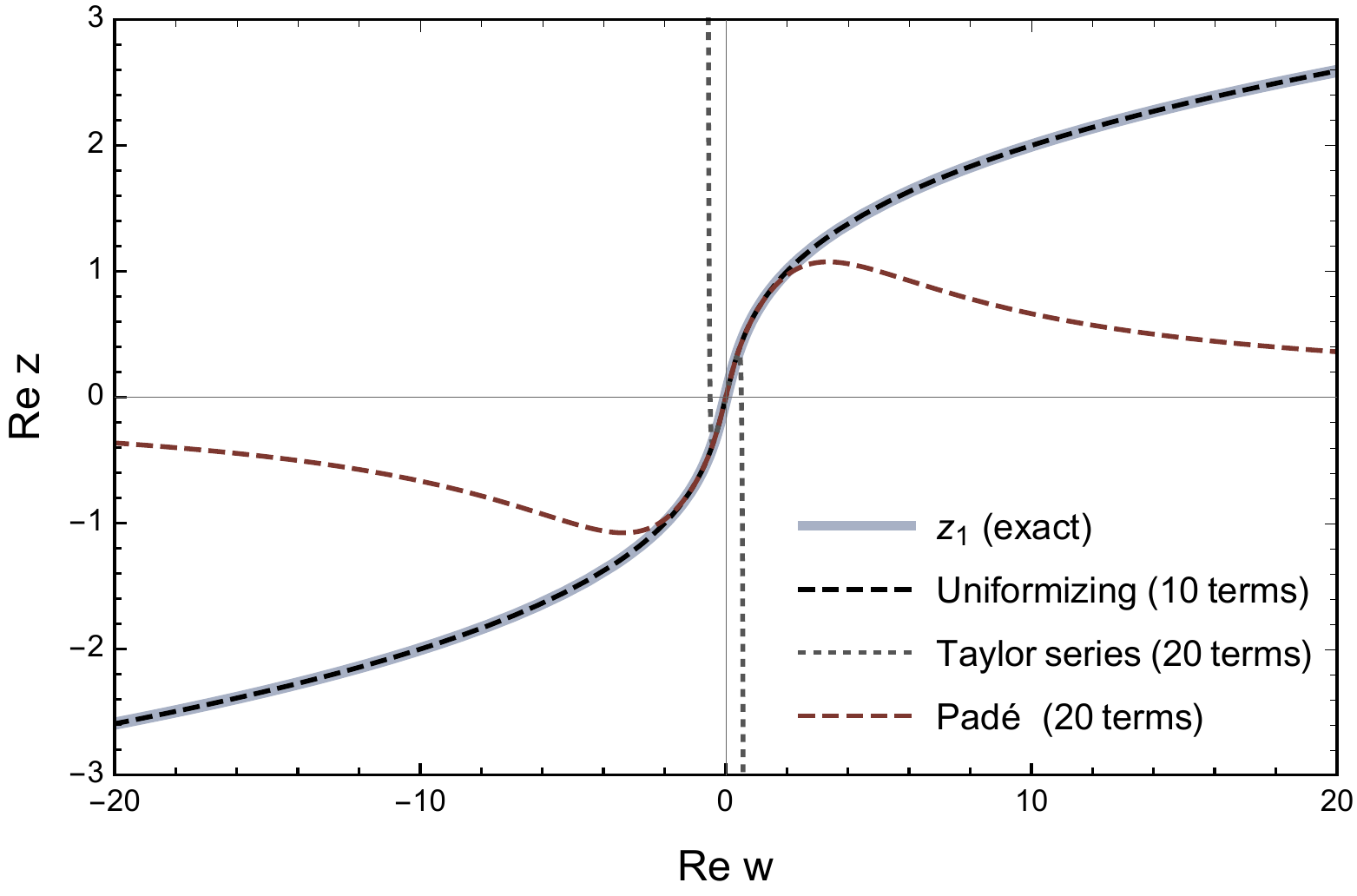}
\caption{The scaled equation of state $z(w)$ for $T>T_c$, shown as the analytic continuation of finite truncations of the Taylor expansion for $z_1(w)$ on the first sheet in (\ref{eq:highT-taylor}). The truncated Taylor series [dotted line] is limited by the radius of convergence, which is determined by the Lee-Yang singularities: $|w_{LY}|\approx 0.385$. A Pad\'e approximation [red dashed line] extends accurately for some distance beyond the radius of convergence, but with the uniformizing map [black dashed line] in \eqref{eq:unimap} the agreement with the exact expression is dramatically more accurate, extending much further. Furthermore, this uniformized analytic continuation was generated with half the number of terms of the initial expansion.}
\label{fig:resum-compare}
\end{figure}

Let us now suppose that we only have access to a finite number of terms of the Taylor series expansion of the equation of state around $h=0$, for various fixed values of $T>T_c$. This corresponds to having a finite number of terms of the Taylor series expansion of $z_1(w)$ around $w=0$: 
\bea
z_1(w)=w - w^3 + 3 w^5 - 12 w^7 + 55 w^9+\dots
\label{eq:highT-taylor}
\ea
This series has a radius of convergence $|w_{LY}|=\frac{2}{3\sqrt{3}}\approx 0.385$, determined by the nearest singularities to the origin on the first sheet, which are the Lee-Yang singularities. Therefore an approximation to the equation of state with a truncated Taylor series can only capture a limited range $w<|w_{LY}|$, regardless of how many terms we have in the expansion. To continue beyond the radius of convergence we can make a Pad\'e approximant of the truncated Taylor series. As shown in Figure \ref{fig:resum-compare}, this leads to an improvement in the direction of $\re w$, along which there are no singularities (the Lee-Yang singularities lie on the $\im w$ line).

However, Pad\'e breaks down along the imaginary $w$ direction because it places poles along the imaginary axis in an attempt to represent the branch cuts. Recall that the Pad\'e approximant is a {\it rational} function so its only singularities are poles. Pad\'e represents a branch cut as an arc of interlacing poles and zeros accumulating to the associated branch point \cite{Stahl,Costin:2020pcj,Costin:2021bay}. This has the consequence that the Pad\'e approximant is not accurate near a cut, and thus does not accurately describe the analytic continuation across a cut. Therefore, the Pad\'e approximant to the truncated equation of state $z=z(w)$, which was developed on the $T>T_c$ sheet, cannot be accurately continued to the $T<T_c$ sheets. 

\begin{figure}[h]
\center
\includegraphics[scale=0.42]{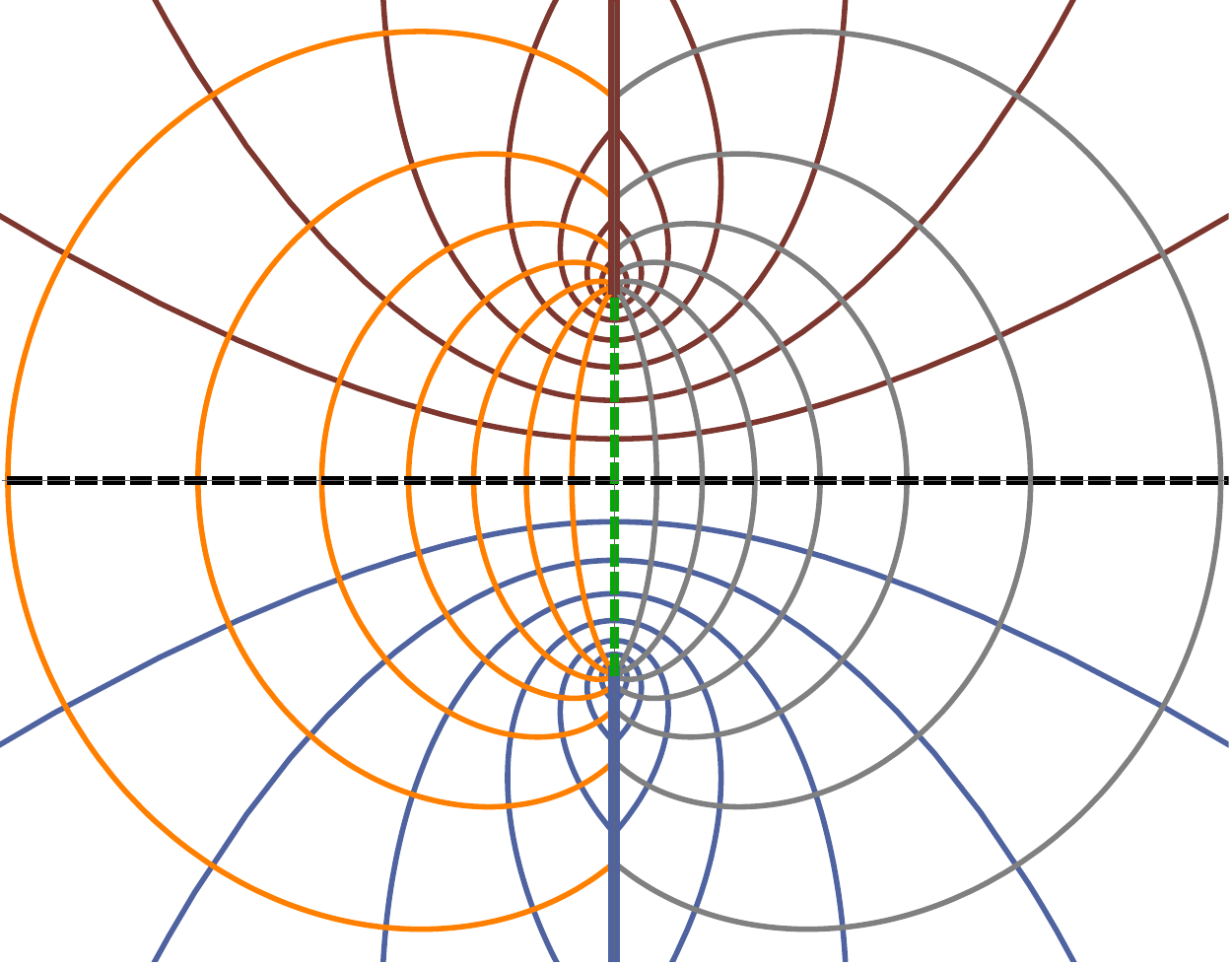}
\includegraphics[scale=0.42]{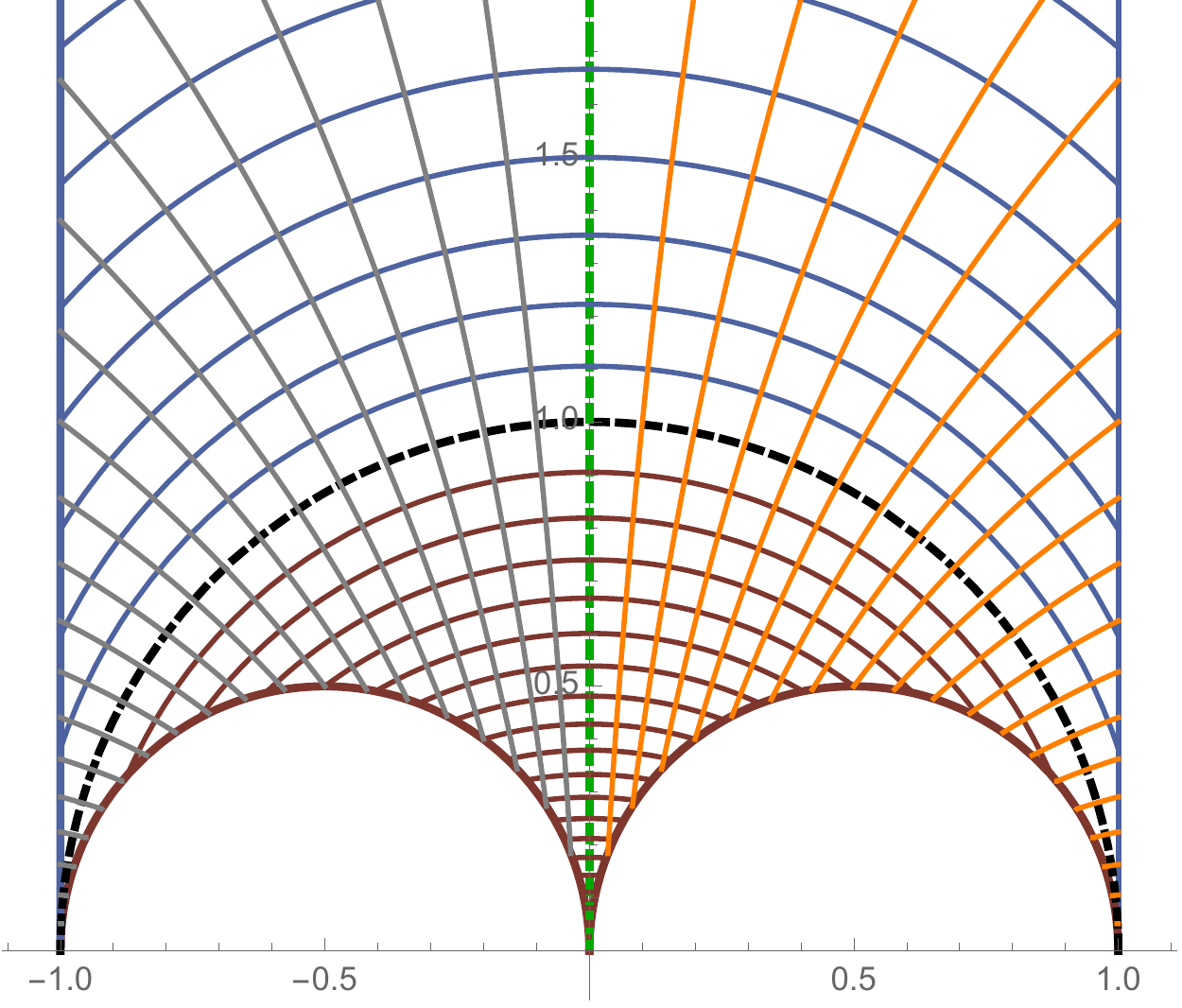}
\includegraphics[scale=0.4]{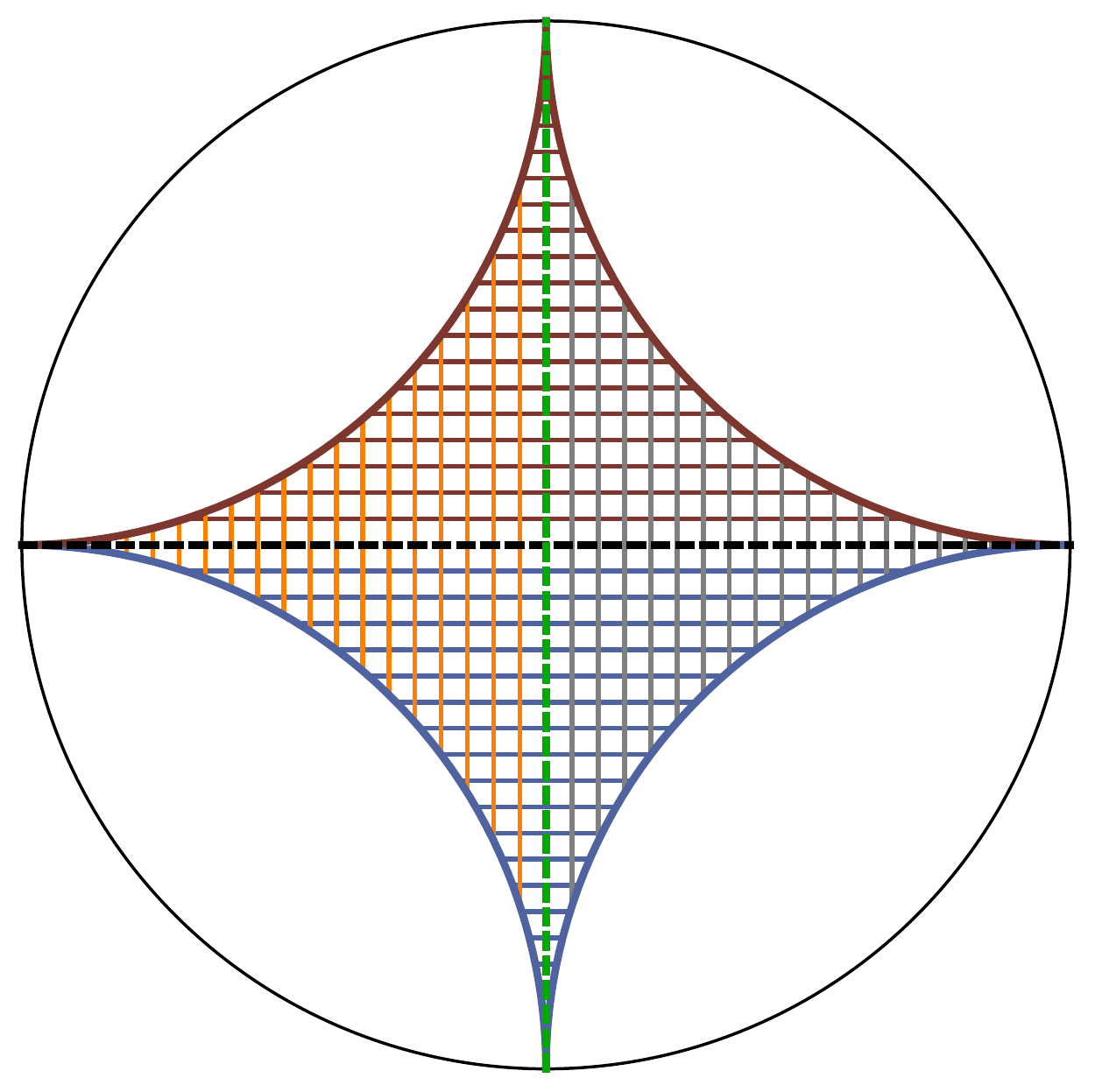}
\caption{Left: The $\tilde w$ plane for the first sheet 
(the high $T$ sheet, $T>T_c$). The red and and blue (upper/lower) lines denote the branch cuts emanating from the LY singularities at $\tilde w=\pm i$. Center: The modular $\tau$ plane after the map $\tilde w\to \tau$ in \eqref{eq:modular}. Right: The unit disk (the $\zeta$ plane) after the map $\tau\to\zeta$ in \eqref{eq:tau-zeta}. The curves with different colors represent the mapping between the $w$,$\tau$ and $\zeta$ planes. }
\label{fig:first-sheet}
\end{figure}

\begin{figure}[h]
\center
\includegraphics[scale=0.42]{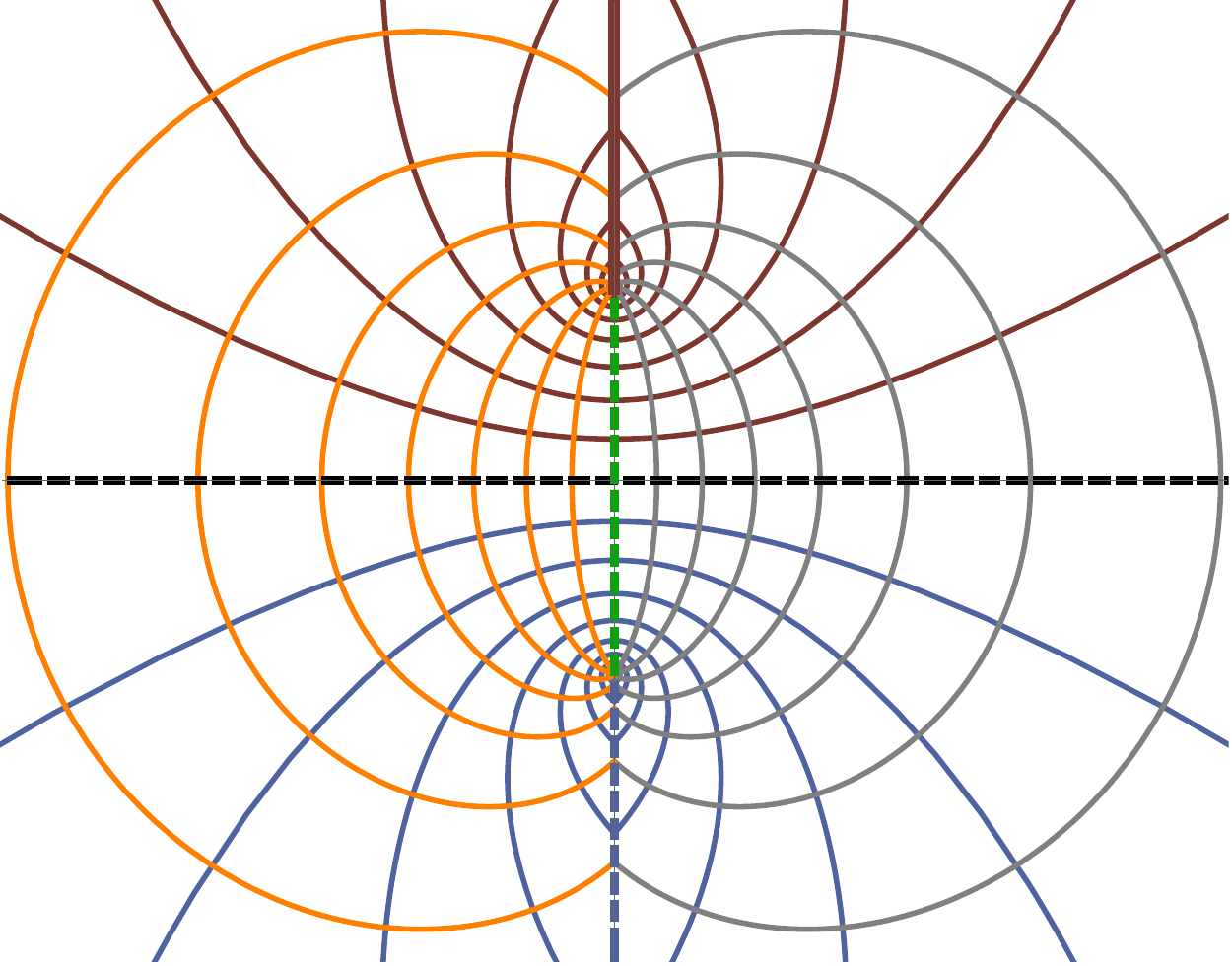}
\includegraphics[scale=0.42]{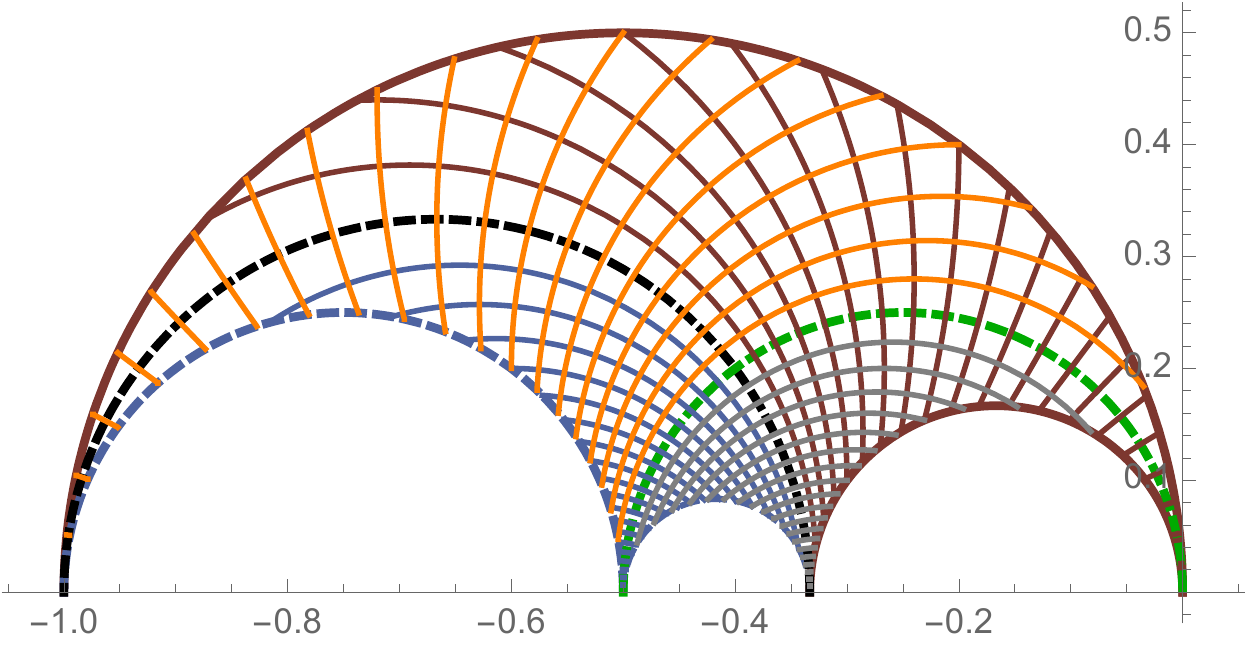}
\includegraphics[scale=0.4]{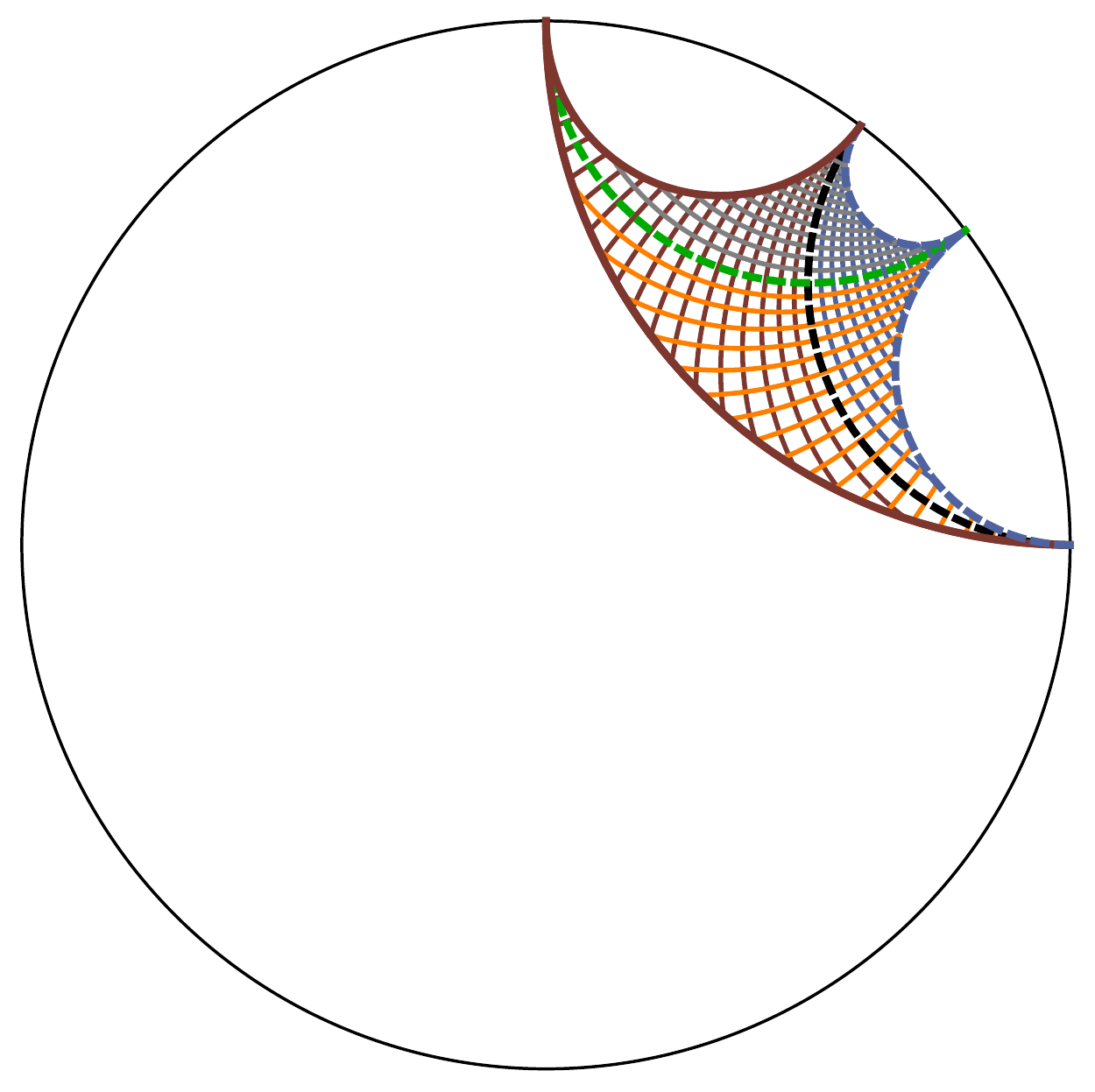}
\caption{The representations of the second sheet 
(the low $T$ sheet, $T<T_c$) in the $\tilde w$ plane (left), in the modular $\tau$ plane (center), and in the $\zeta$ unit disk (right). Note that the low $T$ sheet maps to regions in $\tau$ and $\zeta$ that connect directly to regions associated with the first sheet, shown in Figure \ref{fig:first-sheet}.}
\label{fig:second-sheet}
\end{figure}

This inherent deficiency of Pad\'e can be overcome by {\it first} using a uniformization map, and {\it then} making a Pad\'e approximation \cite{Costin:2020pcj,Costin:2021bay}. For the three-sheeted cubic in \eqref{eq:eos_zw} this can be done exactly, using the following basic facts: 
\begin{enumerate}
\item The equation of state \eqref{eq:eos_zw} is naturally solved in terms of hypergeometric functions (see \eqref{eq:eos_2f1_z1}-\eqref{eq:eos_2f1_lin_z2}), whose analytic continuation properties are well-defined and simple.
\item
The hypergeometric functions are uniformized by an explicit mapping to the upper half plane (see \eqref{eq:modular}), which can then be mapped into the unit disk (see \eqref{eq:tau-zeta}). The resulting uniformization \eqref{eq:unimap} then covers all sheets, which can accessed simply by moving around in the disk.  See Figures \ref{fig:first-sheet},\ref{fig:second-sheet}, and \cite{costin_dynamic} for 
an interactive realization.
\end{enumerate}
The physical implication is two-fold. First, the resulting analytic continuation, starting with exactly the same truncated expansion, is dramatically more accurate, especially near the singularities and cuts. Second, different sheets in the original variable are mapped to different regions in the uniformizing upper half plane, but the result is analytic in the entire upper half plane (or in the entire unit disk), and therefore can be analytically continued between sheets. To make this construction explicit, we first note that the equation of state \eqref{eq:eos_zw} is solved by

\bea
z_1(w)&=&w\,_2F_1\left(\frac23,\frac13,\frac32;-\frac{27w^2}{4}\right) 
\label{eq:eos_2f1_z1}\\
z_2(w)&=&-\frac{w}{2}\,_2F_1\left(\frac23,\frac13,\frac32;-\frac{27w^2}{4}\right)+i\,_2F_1\left(\frac16,-\frac16,\frac12;-\frac{27w^2}{4}\right)\,.
\label{eq:eos_2f1_z2}
\ea
and we recall that $z_3(w)=-z_2(-w)$. 
Furthermore, we can linearize the argument using standard hypergeometric identities (see 15.8.27 and 15.8.28 in \cite{DLMF}):
\bea
z_1(w)&=&-\frac{2 i}{\sqrt3}\left[\,_2F_1\left(\frac13,-\frac13,\frac12;\frac12\left(1-i\tw\right)\right)-
\,_2F_1\left(\frac13,-\frac13,\frac12;\frac12\left(1+i\tw\right)\right)\right]
\label{eq:eos_2f1_lin_z1}\\
z_2(w)&=&\frac{2 i}{\sqrt3}\,_2F_1\left(\frac13,-\frac13,\frac12;\frac12\left(1-i\tw\right)\right)\,.
\label{eq:eos_2f1_lin_z2}
\ea
Here we define the rescaled variable $\tilde w:= \frac{3\sqrt{3}}{2} w$, in terms of which the Lee-Yang singularities are normalized to be at $\tilde{w}=\pm i$. Expressed in this form, all the branch cut technicalities are greatly simplified because the hypergeometric functions have simple connection formulae across a cut \cite{bateman}.

The next step is the crucial one. We use the fact that hypergeometric functions are uniformized by the elliptic modular function $\lambda(\tau)$, where $\tau$ lives in the upper half plane $\im \tau > 0$ \cite{bateman}. This is implemented by the following transformation (to simplify the formulas we use the rescaled variable $\tilde w:= \frac{3\sqrt{3}}{2} w$):
\bea
\tilde w(\tau)= i(-1+2\lambda(\tau)) 
\quad \text{with inverse} \quad
\tau(\tilde w) = i\frac{\mathbb K\left(\frac{1+i \tilde w}{2}\right)}{\mathbb K\left(\frac{1-i \tilde w}{2}\right)}
\label{eq:modular}
\ea
Here $\lambda(\tau)$ is the modular lambda function $\lambda(\tau)=\theta_2^4(\tau)/\theta_3^4(\tau)$, where $\theta_2(\tau)$ and $\theta_3(\tau)$ are the Jacobi elliptic functions: $\theta_2(\tau)=\sum_{n=\infty}^\infty e^{2\pi i \tau (n+1/2)^2}$ and $\theta_3(\tau)=\sum_{n=\infty}^\infty e^{2\pi i \tau n^2}$, defined in the upper half plane $\im \tau>0$. The function $\mathbb K(m)$ in \eqref{eq:modular} is the complete elliptic integral of the first kind:
\bea
\K(m)=\int_0^{\pi/2} \frac{d\theta}{\sqrt{1-m\sin^2\theta}}
\label{eq:k}
\ea
The functions $\lambda(\tau)$ and  $\K(m)$ are implemented in Mathematica as ModularLambda and EllipticK.

It is convenient for both numerical and visualization purposes to combine this map with a subsequent map that takes the upper half $\tau$ plane into the unit disk $|\zeta|<1$:
\bea
\tau(\zeta):=i\left(\frac{1+i\zeta}{1-i\zeta}\right)
\quad \text{with inverse} \quad
\zeta=i\left( \frac{1+i\tau}{1-i \tau}\right)
\label{eq:tau-zeta}
\ea
The combined transformation, from the three-sheeted $w$ plane directly to the unit disk in $\zeta$ is:
\bea
\tilde w=i \left[-1+2\lambda\left(i\frac{1+i\zeta}{1-i\zeta} \right) \right] 
\quad \text{with inverse} \quad
\zeta =i\frac
{\K \prt{\frac{1-i \tilde w}2}-\K\prt{\frac{1+i \tilde w}2}}
{\K\prt{\frac{1-i \tilde w}2}+\K\prt{\frac{1+i \tilde w}2}}
\label{eq:unimap}
\ea
The geometrical properties of these maps are depicted in Figures \ref{fig:first-sheet}, \ref{fig:second-sheet} and \ref{fig:unimap}. Figure \ref{fig:first-sheet} shows how the first sheet in the $\tilde w$ variable is mapped to a specific region of the upper half complex $\tau$ plane, and subsequently to a specific "circular quadrilateral" region of the unit disk in the $\zeta$ variable (right-hand plot in Figures \ref{fig:first-sheet}). Figure \ref{fig:second-sheet} shows how the second sheet in the $\tilde w$ variable is mapped to a different but connected portion of the upper half $\tau$ plane, and correspondingly to a different but connected portion of the unit disk in the $\zeta$ variable. These two regions are related by a modular transformation. This can be continued {\it ad infinitum}, with the trajectories in the unit disk in $\zeta$ encoding all possible trajectories on the three-sheeted Riemann surface in the $\tilde w$ variable. Examples are given below, and see \cite{costin_dynamic} for 
an interactive realization.

\subsection{Approximate Reconstruction of the Ising Equation of State: the Uniformized-Pad\'e Method}
\label{sec:optimal}

The maps described in the previous section give an {\it exact} uniformization of $z(w)$ on its entire Riemann surface. However, the main practical use of these maps is as building blocks for {\it approximate} uniformizing maps. For example, suppose we know (or conjecture) that a finite-order expansion about the origin is limited by two dominant singularities, which are in general branch points.\footnote{This is a common occurrence in physical applications: e.g. \cite{Caliceti:2007ra,Florkowski:2017olj,Rossi:2018urw,Serone:2019szm,Aniceto:2018uik,bertand}.} Then an approximate uniformizing map can be used, based solely on this (conjectured) information. If the two branch points are symmetrically located, then the map is precisely \eqref{eq:unimap}, but the corresponding map is known when these two dominant singularities have general locations \cite{nehari,kober,Costin:2020pcj,Costin:2021bay}.

Now suppose we do not have the exact solutions $z_1(w)$ and $z_2(w)$ of the scaled equation of state \eqref{eq:eos_zw}, but just a finite-order truncated expansion for $z_1(w)$, generated in the high temperature region. We first locate the singularities that limit the convergence: this can be done approximately, using for example methods described in Section \ref{sec:conf_pade} (or even more accurate methods in \cite{Costin:2020pcj,Costin:2021bay}). We learn that there are two dominant singularities. Let us rescale the variable $w$ to place these at $\tilde w=\pm i$. 

The actual implementation of the reconstruction method is extremely simple. We note that the map from $\tilde w$ to $\zeta$ in \eqref{eq:unimap} takes the origin $\tilde w=0$ to the origin of the disk, $\zeta=0$: 
\bea
\tilde w(\zeta)&=& i \left[-1+2\lambda\left(i\frac{1+i\zeta}{1-i\zeta} \right) \right] \nn
&=& \frac{\pi^2\zeta}{\Gamma\left(\frac{3}{4}\right)^4}
+
\frac{1}{4}\left(\frac{\pi^2\zeta}{\Gamma\left(\frac{3}{4}\right)^4}\right)^3 +\frac{13}{240}\left(\frac{\pi^2\zeta}{\Gamma\left(\frac{3}{4}\right)^4}\right)^5 +\dots ... 
\label{eq:w-zeta-taylor}
\ea
This means that we simply compose the series expansions to convert the finite-order expansion in $\tilde w$ into a finite-order expansion in $\zeta$. We then Pad\'e in $\zeta$ and map back to $\tilde w$. The resulting Uniformized-Pad\'e procedure is \cite{Costin:2020pcj,Costin:2021bay}:
\begin{enumerate}
  \item 
  Re-expand the original truncated Taylor series in $\tilde w$ as a Taylor series in $\zeta$, and truncate at the same order in $\zeta$. This procedure is optimal \cite{Costin:2020pcj}.
  \item
  Make a Pad\'e approximant of the resulting truncated series in terms of $\zeta$.
  \item
  Map this Pad\'e approximant back to the $\tilde w$ plane using the inverse map in \eqref{eq:unimap}. (Note that in contrast with the case of the conformal map \eqref{eq:conf_map}, here the uniformizing map and its inverse are both explicit.)
\end{enumerate}
The output of this simple algorithm is an analytic continuation of the the truncated expansion of $z_1(w)$, which is of high precision on the first (high temperature) sheet, and which smoothly crosses to other sheets, in particular to the low temperature region. The high precision also means that the procedure can be iterated to refine an initial estimate of the singularity locations.

\subsubsection{Improved accuracy on the high temperature sheet}

To quantify the high precision of this Uniformized-Pad\'e analytic continuation on the first ($T>T_c$) sheet, we compare it in Figure \ref{fig:resum-compare} with the exact expression for $z_1(w)$ in \eqref{eq:eos_2f1_z1} (or \eqref{eq:eos_2f1_lin_z1}), with the Taylor expansion of $z_1(w)$ truncated after 20 terms, and with the (near diagonal) Pad\'e approximant of this 20-term truncated expansion. As expected, the truncated series breaks down at the Lee-Yang radius of convergence $\frac{2}{3\sqrt{3}}\approx 0.385$. The Pad\'e approximant is better, going beyond the radius of convergence, but it breaks down at $\re w\approx \pm 2$. By comparison, the Uniformized-Pad\'e approximation matches very accurately the exact result for $z_1(w)$ all the way out to $\re w=\pm 20$, well beyond the Lee-Yang radius of convergence. Furthermore, this Uniformized-Pad\'e approximation was implemented using half the number of input terms. This higher precision shown in Figure \ref{fig:resum-compare} is along the $\re w$ axis, which does not encounter any singularities or cross any cuts. The improvement is even more dramatic along the imaginary $w$ axis, where Pad\'e fails already at the radius of convergence, where it first encounters the branch points.

Note that the construction of the uniformizing map in \eqref{eq:unimap} uses knowledge of the location of the Lee-Yang singularities. As discussed in Section \ref{sec:conf_pade}, if these locations are unknown they can be found numerically to high precision by iteration \cite{Costin:2020pcj,Costin:2021bay}. The fact that the uniformizing map \eqref{eq:unimap} leads to an {\it exact} uniformization of the function relies on the underlying Riemann surface being that associated with the hypergeometric functions, which solve the equation of state \eqref{eq:eos_zw}. However, even if it were not the exact Riemann surface, the use of this uniformizing map produces significantly higher precision than other methods for any problem with a pair of leading singularities \cite{Costin:2020pcj,Costin:2021bay}, which is a common occurrence in a wide range of physical applications.

\subsubsection{Analytic continuation from the high temperature sheet to the low temperature sheets}

\begin{figure}[h]
\center
\includegraphics[scale=0.5]{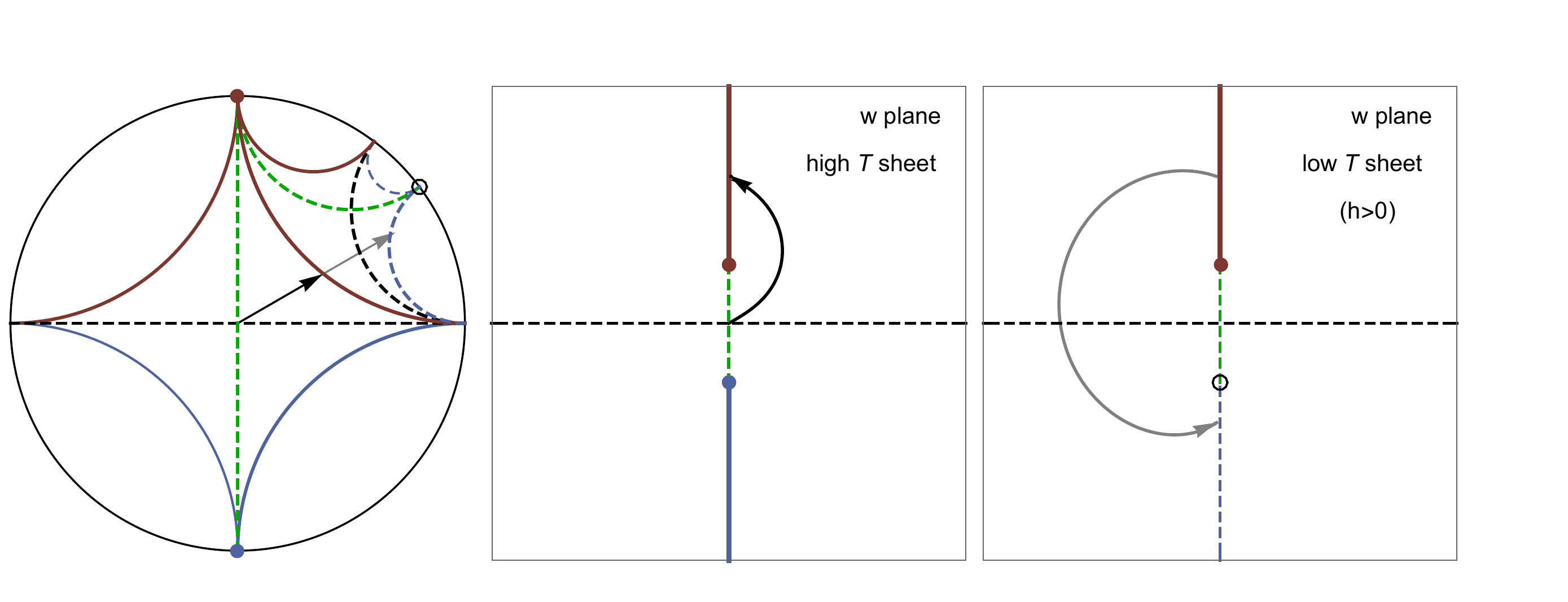}
\includegraphics[scale=0.5]{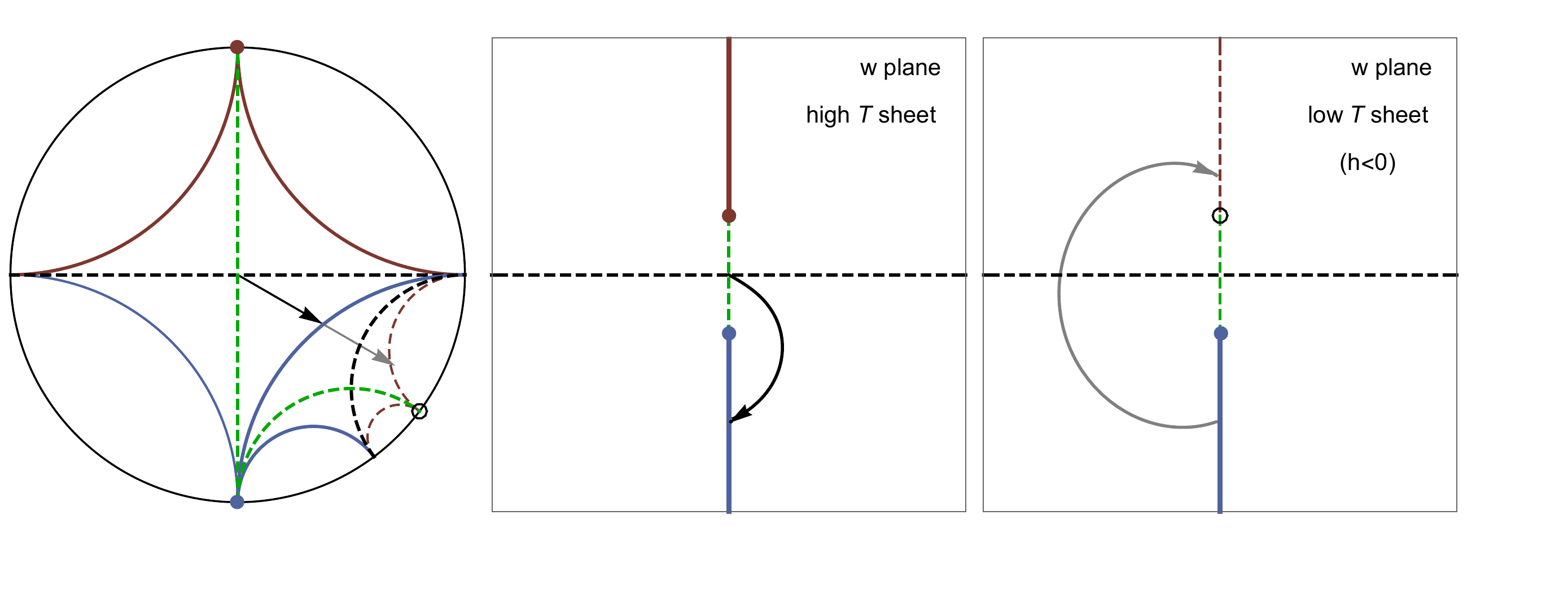}
\caption{The analytic continuation from the high $T$ sheet to the low $T$ sheets with $h>0$ (upper) and $h<0$ (lower). The arrows showing trajectories inside the unit disk match the trajectories that cross from the high $T$ sheet to the low $T$ sheet in the $w$ variable. The corresponding boundaries of the sheets (i.e. the edges of the cuts) become circular boundaries inside the unit disk, obtained by Schwarz reflections. }
\label{fig:unimap}
\end{figure}
In addition to improved accuracy on the first sheet, a distinguishing feature of the Uniformized-Pad\'e approximation is its ability to reconstruct the underlying function globally, making it possible to pass to higher Riemann sheets, even when starting from a finite-order truncated approximation \cite{Costin:2020pcj,Costin:2021bay}.
This can be seen already in the behavior of the approximation along the imaginary $w$ axis, which encounters the Lee-Yang singularities at $\tilde w=\pm i$, and the associated branch cuts that need to be crossed in order to pass from the high temperature sheet to the low temperature sheets.

The transition from one sheet to another works as follows. Consider the physical problem of trying to evaluate the low temperature equation of state starting from a truncated expansion of the high temperature equation of state. This (truncated) high temperature expansion is generated on the first sheet, but we want the solution on the second (low $T$) sheet. 
\begin{figure}[h]
\center
\includegraphics[scale=0.5]{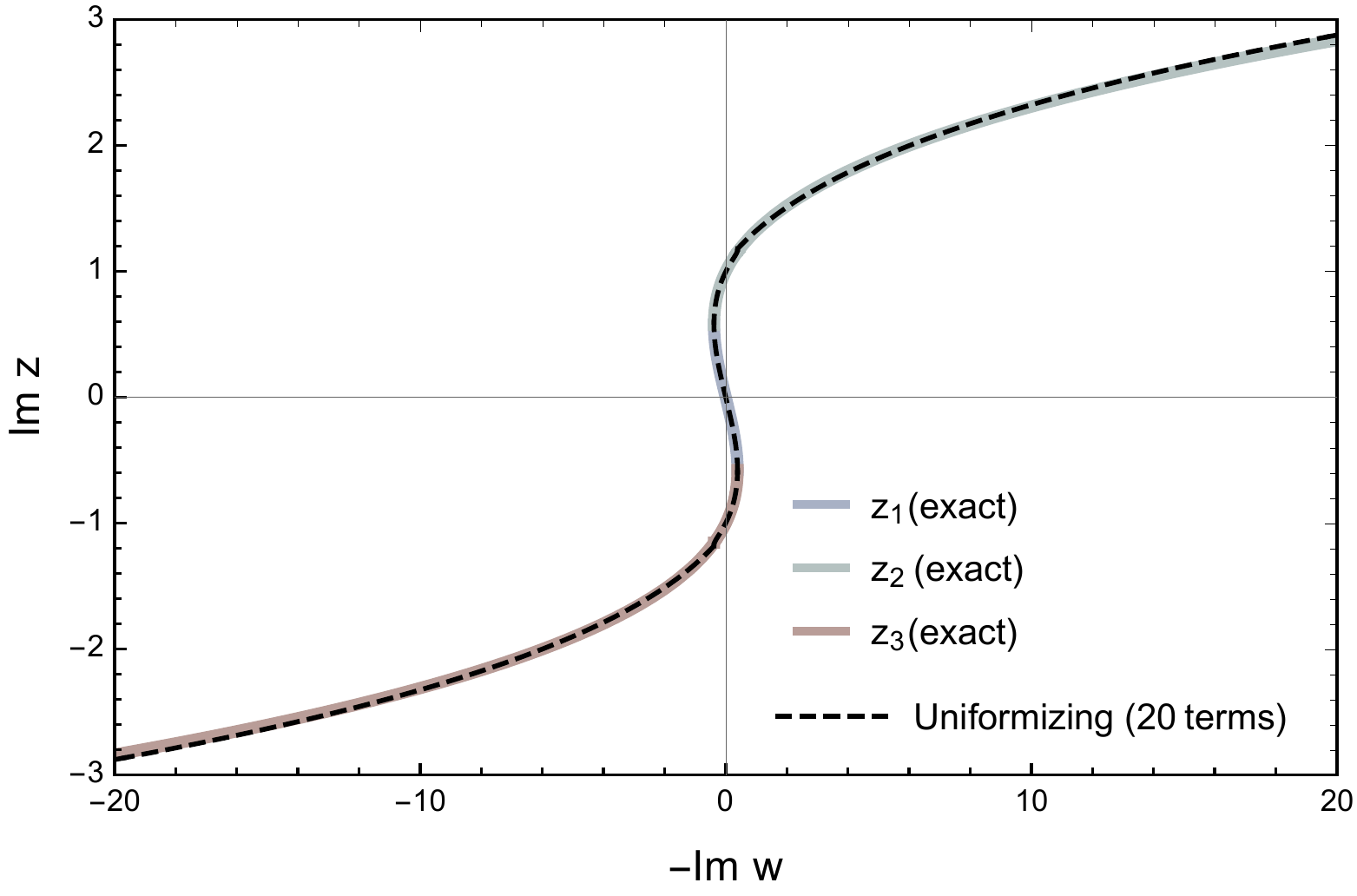}
\includegraphics[scale=0.5]{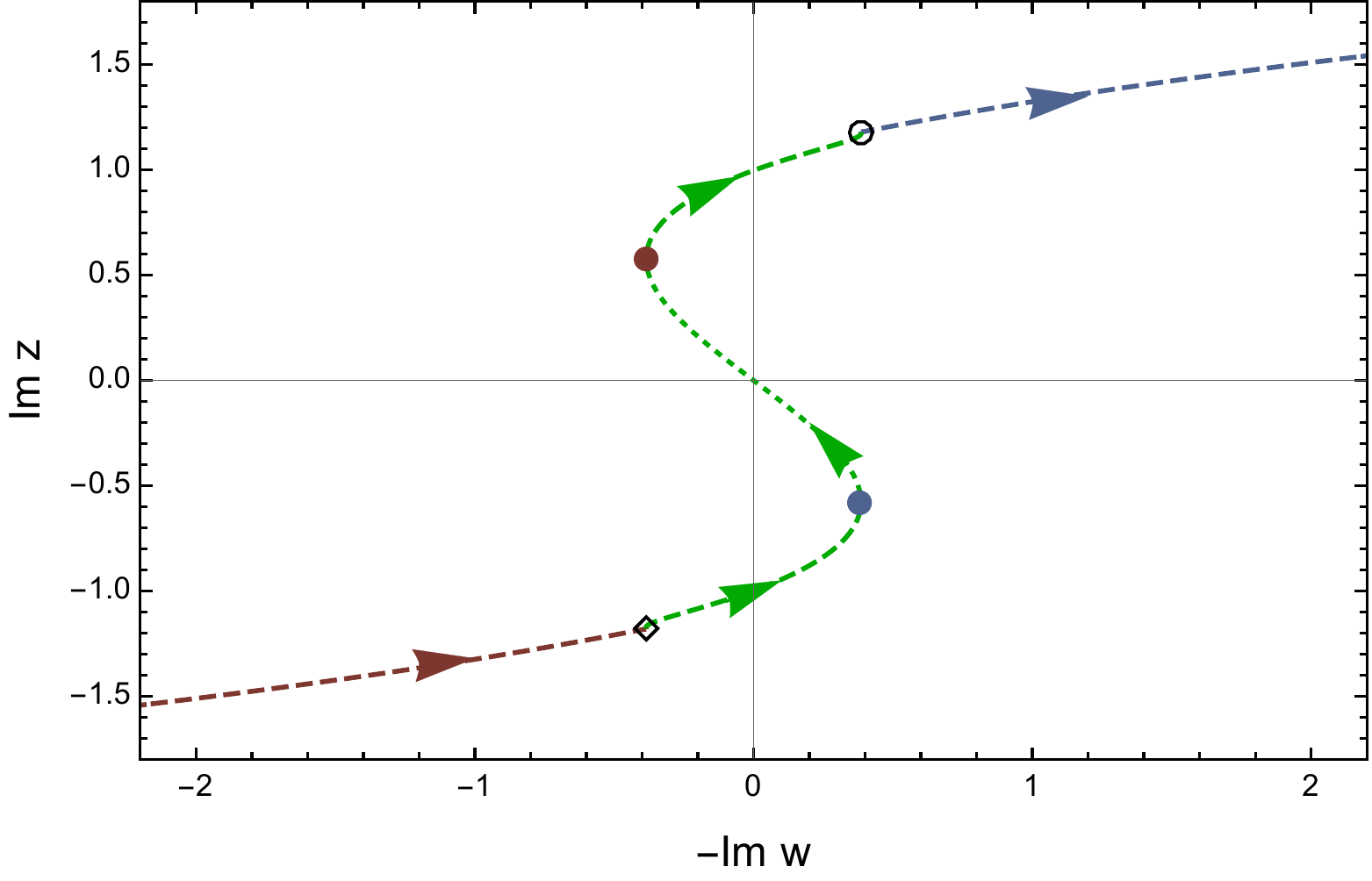}
\caption{The low $T$ equation of state,  in terms of the scaled variables $z$ and $w$, reconstructed from the high $T$ expansion using uniformed Pad\'e, compared with the exact result (left). Note that with just 20 input terms for the expansion of $z_1(w)$ on the high $T$ sheet we can reconstruct the solution on other sheets with high precision. The right-hand panel shows a zoomed-in view, highlighting the trajectory in the $w$ plane and in the unit disk, as shown in Fig. \ref{fig:resum-traj}. }
\label{fig:resum-uniform}
\end{figure}
\begin{figure}[h]
\center
\includegraphics[scale=0.55]{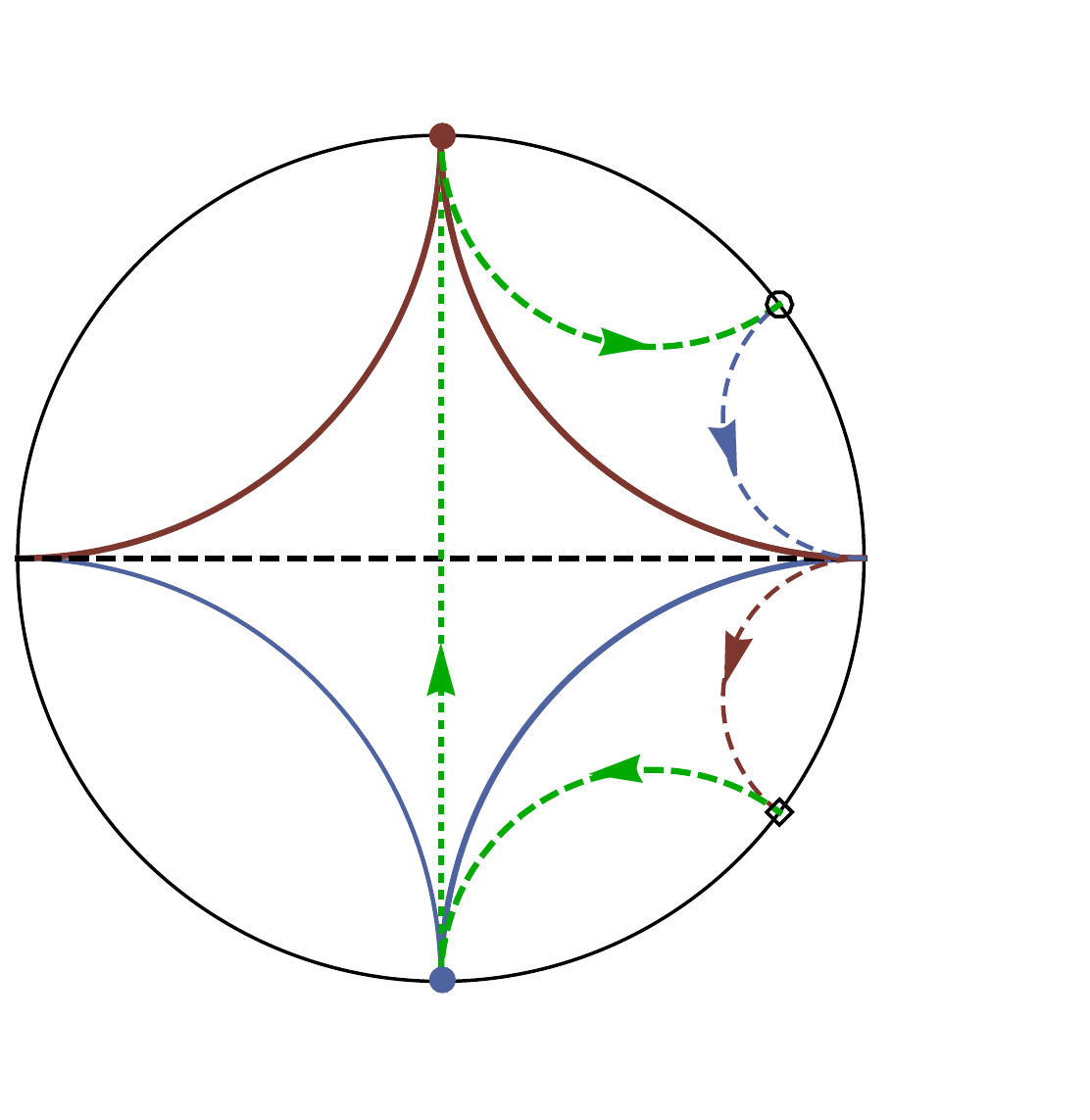}
\includegraphics[scale=0.52]{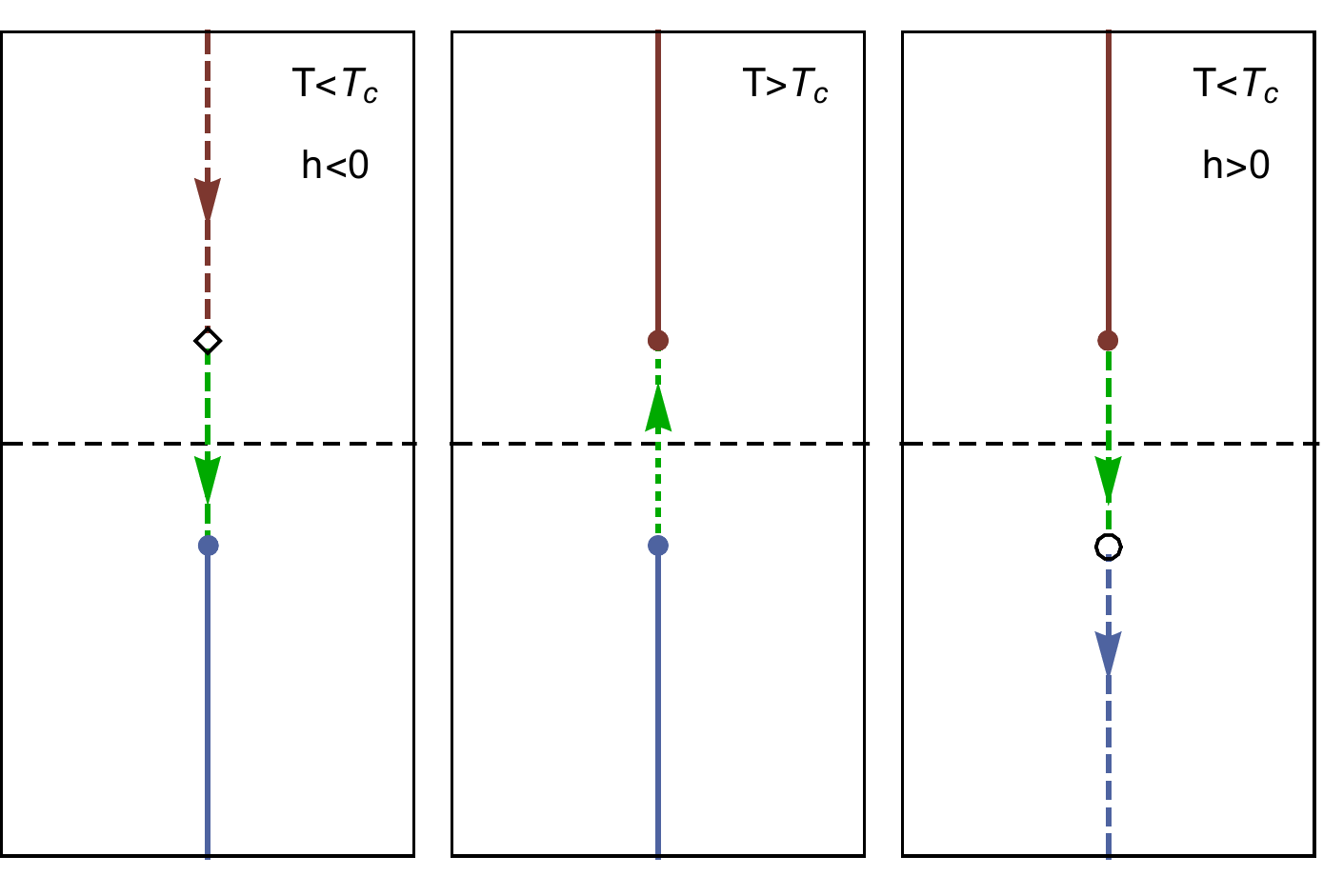}
\caption{The trajectory that captures the low $T$ equation of state $M(h)$ in the scaling variables in the unit disk (left) and in the $w$ plane (right). The trajectory goes through three different Riemann sheets (right) in the $w$ plane but is smooth and entirely contained in the unit disk (left). }
\label{fig:resum-traj}
\end{figure}
As shown in Figure \ref{fig:first-sheet}, the uniformizing map takes the first sheet to a circular quadrilateral inside the unit disk, for the $\zeta$ variable. Suppose we cross to the low $T$ branch by crossing the upper branch cut (shown in red). The low $T$ sheet in this case is represented in the unit disk as a region obtained by performing a Schwartz reflection with respect to the image of the branch cut in the unit circle i.e. the upper right red circle in Fig. \ref{fig:first-sheet} (right). The image of the low $T$ sheet obtained this way is shown in Fig. \ref{fig:second-sheet} (right). In the modular plane this corresponds to a particular Mobius transformation which maps the image of the first sheet in Fig. \ref{fig:first-sheet} (center) inside the region bounded by the the semicircle (see Fig. \ref{fig:second-sheet} (center)). Notice that in this low $T$ sheet (where $h<0$), there is only one branch cut, shown in solid red. As opposed to the high $T$ sheet, the line segment $\im w< -2/(3\sqrt{3})$ is not a branch cut and is therefore shown as a dashed blue line in Fig. \ref{fig:second-sheet}. In fact this is the region where the physical equation of state for $T<T_c$ and $h>0$ is defined, namely the magnetization $M\propto-\im w$ with $\im w<-2/(3\sqrt{3})$. 

This is illustrated in Figure \ref{fig:unimap}, which shows how the continuation from the high temperature sheet to the low temperature sheet appears in the unit disk. An illustrative trajectory in the $w$ plane that goes from the high $T$ branch to the low $T$ branch and its image on the unit disk are shown in Fig. \ref{fig:unimap}. This trajectory, in the $w$ plane, starts from the origin, crosses the upper branch cut to the low $T$ sheet and finally ends at some point on the negative imaginary axis with $\im w<-2/3\sqrt{3}$, which is proportional to some value of the magnetization with $T<T_c$ and $h>0$ in the stable branch of the equation of state. In the unit disk, this trajectory is a line segment shown in Fig. \ref{fig:unimap} (top). The main point we emphasize is that even though in the $w$ plane the trajectory goes through different sheets, in the unit disk it is completely regular. Similarly, it is also possible to analytically continue to the $h<0$ sheet by going through the lower branch cut (blue line in Fig. \ref{fig:first-sheet}, left) as shown in Fig. \ref{fig:unimap} (bottom).

Figures \ref{fig:resum-uniform} and \ref{fig:resum-traj} show how the low temperature equation of state can be reconstructed from just 20 input terms 
of the Taylor expansion of the high temperature equation of state.
In Fig. \ref{fig:resum-uniform} (left) we compare the exact result with the result obtained from the uniformized-Pad\'e approximation. Similar to the high $T$ expansion in Figure \ref{fig:resum-compare}, the numerical accuracy in the low T region is also remarkable and extends to large values $|w|\approx20$. We stress that this result is obtained from the Taylor expansion of $z_1(w)$ for $T>T_c$ and analytically continued via the uniformizing map. It is not the expansion of $z_2(w)$ or $z_3(w)$. Nevertheless it captures the low $T$ behavior of the equation of state remarkably well in a region where neither the truncated Taylor series nor the Pad\'e approximant has any applicability whatsoever.

\section{Summary and Conclusions}
\label{sec:conclusions}

In this paper we described a robust framework to reconstruct efficiently the equation of state of a thermodynamic system near a critical point $T_c,\mu_c$, using only a finite number of coefficients from a local expansion at $\mu=0$. We first showed that pairing the usual Pad\'e resummation with a conformal map significantly improves the approximation to the underlying equation of state, compared to simply summing the truncated Taylor series or performing a regular Pad\'e approximation. The most important improvement is the extension of the range of the approximation. The applicability of the Taylor series is limited by its radius of convergence. Ordinary Pad\'e resummation allows one to go pass beyond the radius of convergence but produces unphysical singularities which limit the improvement in physically important regions. Conformal Pad\'e eliminates these unphysical singularities which leads to a dramatic improvement in the range of the approximation. In particular we showed (see Figure \ref{fig:pade-chis}) that it captures the characteristic features of the susceptibilities in the vicinity of the critical point which play a significant role in the search for the QCD critical point.

We also showed that it is even possible to analytically continue to higher Riemann sheets by pairing Pad\'e resummation with a suitably engineered map, namely a uniformizing map. We demonstrated this procedure in the Ising model. Physically, this makes it possible to analytically continue an expansion obtained in the high $T$ crossover region ($r>0$) to the low $T$ first-order region ($r<0$). The only input we needed was the Taylor coefficients and the location of the Lee-Yang singularity whose value can be approximated by the same iteration procedure explained in Sec. \ref{sec:conf_pade}. 

There are various future directions left for future work. An important extension is to go beyond the mean field limit. For the Ising model, in the crossover region ($r>0$), which corresponds to the first sheet, the two-cut nature of the $w$ plane is universal \cite{Lee:1952ig} albeit with different branch point singularities determined by the critical exponents, $\bd$. Given that the $r<0$ region has more structure, such as the Langer cut \cite{An:2016lni,An:2017brc,An:2017rfa}, it would be interesting to extend this machinery beyond the mean field. Beyond the mean field approximation, even though in general the equation of state does not have an analytic representation, knowledge of the universality class and/or the critical exponents can help to construct an {\it approximate} 
uniformizing map that enables improved analytic continuation. Another interesting aspect of beyond the mean field case is that the equation of state, $w=F(z)$, can be expressed as an $\epsilon$ expansion which is asymptotic. Related conformal Pad\'e tecnhiques are well known in the study of the $\epsilon$ expansion \cite{Guida:1998bx,ZinnJustin:2002ru}. 
One could construct a hybrid resummation scheme that involves uniformizing and conformal Pad\'e both in $z$ and the Borel plane of $\epsilon$. Alternatively one could used the parametric representation of the equation of state \cite{Wallace_1974}. Another possible direction is to incorporate the analytical continuation scheme introduced in this paper with the expansions obtained with pure imaginary chemical potential, as for QCD it is very challenging to compute beyond the first few terms of the Taylor expansion with real $\mu$. 
Finally  it is also necessary to address the issue of noise in the Taylor coefficients as they are typically computed via stochastic methods which unavoidably introduces noise.  It is therefore important to ensure the stability of these resummation methods with noisy data.

 \vspace{.5cm}
\noindent {\bf Acknowledgments} \\
This work is supported in part by the U.S. Department of Energy, Office of High Energy Physics, Award DE-SC0010339 (GD) and the Junior Faculty Development Award form UNC-CH (GB). We thank Ovidiu Costin for discussions.

\bibliographystyle{utphys}
\bibliography{refs}
\end{document}